\documentclass[conference]{IEEEtran}

\usepackage{epsf}
\usepackage{graphicx}
\usepackage{floatflt}
\usepackage{color}
\usepackage{comment}
\usepackage[utf8]{inputenc}
\usepackage{amssymb}
\usepackage{amsmath}
\usepackage{algorithm}
\usepackage{algpseudocode}
\usepackage{cleveref}
\usepackage{cite}

\usepackage{multirow}

\floatname{algorithm}{Procedure}

\newcommand{\specialcell}[2][c]{%
  \begin{tabular}[#1]{@{}c@{}}#2\end{tabular}}

\ifCLASSINFOpdf
\else
\fi
\begin{document}
%

\title{A Survey of Software-Defined Smart Grid Networks: Security Threats and Defense Techniques}

\author{\IEEEauthorblockN{\hspace{0.08in}Dennis Agnew \hspace{0.08in} Sharon Boamah \hspace{0.08in}Janise McNair}
\IEEEauthorblockA{\textit{Department of Electrical and Computer Engineering},
\textit{University of Florida, Gainesville, FL }\\
\{dennisagnew, sharonboamah\}@ufl.edu, mcnair@ece.ufl.edu
}

}


%


\maketitle
\setcounter{page}{1}
\begin{abstract}

Smart grids are replacing conventional power grids due to rising electricity use, failing infrastructure, and reliability problems. Two-way communication, demand-side administration, and real-time pricing make smart grids (SGs) dependent on its communication system. Manual network administration slows down SG communication. SG networks additionally utilize hardware and software from several vendors, allowing devices to communicate. Software-defined SGs (SD-SG) use software-defined networking (SDN) to monitor and regulate SG global communication networks to address these concerns. SDN separates the data plane (routers and switches) from the control plane (routing logic) and centralizes control into the SDN controller. This helps network operators manage visibility, control, and security. These benefits have made SDN popular in SG architectural and security studies. But because SD-SGs are vulnerable to cyberattacks, there are concerns about the security of these SD-SG networks. Cybercriminals can attack software-defined communication networks, affecting the power grid. Unauthorized access can be used to intercept messages and introduce false data into system measurements, flood communication channels with fraudulent data packets, or target controllers, a potential single point of failure, to cripple SDN networks. Current research reflects this paradigm as defense and security against such attacks have developed and evolved. There is a need for a current study that provides a more detailed analysis and description of SD-SG network security dangers and countermeasures, as well as future research needs and developing threats for the sector. To fill this void, this survey is presented.

\end{abstract}

\section{Introduction}
\label{sec_intro}

Smart grids (SGs) are improved versions of traditional power grids and rely on two-way communication technology to improve efficiency, reliability, and sustainability. Unlike the outdated and inefficient traditional power grids, which suffers from frequent power outages, SGs uses a variety of technologies, such as sensors, analytic, and control systems, to more effectively monitor and manage energy consumption, generation, and distribution \cite{rehmani2019software}. SG can monitor energy demand and supply in real-time, making them more adaptable to changing energy needs.
\begin{figure*}[ht]
\centering
\includegraphics[width=1.5\columnwidth]{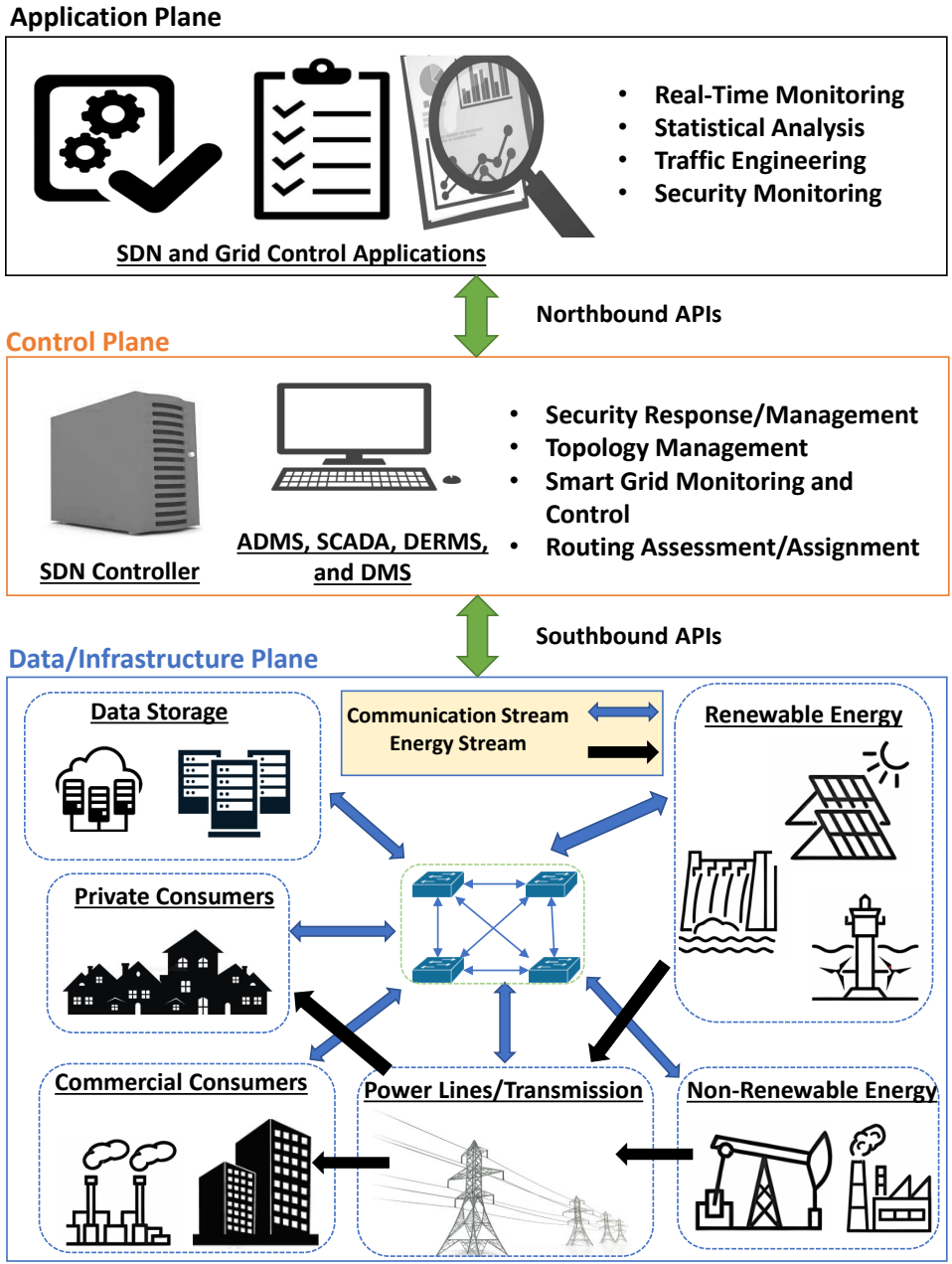}
\caption{SD-SG Architecture}
\label{fig:SDN_based_smart_grid_arch}
\end{figure*}

SG provides greater transparency and accessibility to energy providers, allowing them to effectively monitor and manage their energy consumption in real-time\cite{butt2021recent}. They also allow utility companies to communicate with customers more efficiently, providing real-time feedback on energy consumption and assisting customers in making informed decisions about energy usage \cite{sirojan2019embedded}. These features are due to a high-level interdependence between the power grid level and the networking level of SG architectures~\cite{agnew2022implementation, kong2020routing, aljohani2022cross}. However, high-level of interdependency is also demonstrated during cyberattack scenarios as an attack in the networking layer can alter the behavior in the power grid layer as well\cite{agnew2022implementation}. Thus, network security is a critical component of SG and they become more vulnerable to cyberattacks as they become more interconnected and reliant on communication networks \cite{priyadarshini2021identifying, kong2020review}.

Cyberattacks could have severe consequences, ranging from power outages for customers and millions of dollars of damages for providers~\cite{singh2018entropy}. In addition, SG need continuous monitoring and control and necessitates a network that can handle large amounts of data quickly and reliably, while also ensuring that data is delivered to the correct location and on time \cite{kumari2019fog}. As the SG infrastructure expands, the network's complexity grows, necessitating more sophisticated management tools and expertise to ensure optimal network performance and security \cite{fan2021restoration}. The network must be able to handle varying energy demand and supply, which can change quickly and necessitate real-time network adjustments, and ongoing management to ensure the safe, dependable, and efficient delivery of energy to consumers. To address these issues of security, flexibility, and network management, current research efforts  are investigating the application of software-defined networking (SDN) to improve network security and performance in SG \cite{rehmani2019software, aggarwal2021survey}. 

 Software-Defined Networking (SDN) is a network design technique that allows networks to be intelligently and centrally controlled, or programmed, through the use of software applications. SDN separates the control and data planes while integrating the control plane into the controller. This allows for greater flexibility and scalability in network infrastructure management \cite{maleh2022comprehensive}. Software-defined networking (SDN) research has shown rapid improvements and discoveries since its public launch in 2009. Compared to traditional networks, SDN has better utilization, efficiency of resources, flexibility of network services, and reduced cost of maintenance \cite{sun2020detecting}. Managing the complex infrastructure of a SG can be a daunting task with traditional networking, requiring significant manual intervention and human resources \cite{rehmani2019software}. As shown in Figure~\ref{fig:SDN_based_smart_grid_arch}, SDN simplifies this process by providing a centralized software controller that manages and configures network devices and protocols, automating many of the tasks involved in network management. The adoption of SDN with SG have created software defined SG (SD-SG) which experience improved network security, visibility and control by enabling real-time monitoring and analytics, allowing for proactive identification and resolution of network issues \cite{kabbara2022towards}. This level of visibility and control is critical in managing the dynamic and complex infrastructure of a SG, where energy source availability can change rapidly. 

  Although SDN can provide security response solutions, SD-SG are susceptible to cyberattacks such as distributed denial of service (DDoS), unauthorized access, false data injection, etc \cite{agnew2022implementation}. Furthermore, the introduction of SDN introduces the SD-SG to additional attacks that target  specific elements of the SDN network such as the controller. Current research is disjointed and focuses on development of security frameworks that protect these SD-SG from these attacks by leveraging the features of SDN. There is a need for a review that centralizes, summarizes, and analyzes these approaches for the development of future efforts. Other literature \cite{rehmani2019software,ibdah2017security,akkaya2015software,abujubbeh2019software, demirci2018software,kim2015trends} have surveyed SD-SG and only briefly touches on network security  or are now outdated and do not contain recent techniques and attack types in literature as this survey as summarized in Table~\ref{Related_Works_Table}. As a result, we present this survey to fill this void by providing a comprehensive, up-to-date survey of current threats to SD-SG network security as well as novel future directions and open challenges for SD-SG network security research. Therefore, the contributions of this survey are as follows:
 \begin{itemize}
     \item An updated survey of cyberattacks affecting SD-SG and current SD-SG mitigation techniques, with a focus on novel research efforts in the last 5 years      
     \item A review of open, unresolved cyberattacks that are not addressed in the current literature for SD-SG network security and discussion of emerging threats
     \item A review of potential mitigation techniques for the emerging cyberattacks for SD-SG network security
 \end{itemize}
 
 The rest of the paper is organized as follows. Section~\ref{related works} describes related work and compares them to this survey. Section~\ref{Background} provides background information of SDN and SD-SG. Section~\ref{What are software defined smart grids?} defines software defined SGs and provides a descriptive definition and review. The literature review of SD-SG network security threats and their defense mechanisms  are discussed in section~\ref{Network Security Challenges of Software-Defined Smart Grids}. The unsolved challenges for SD-SG security and emerging cyber threats are discussed in section~\ref{Unsolved Challenges}. Lastly, the paper is concluded in section~\ref{conclusion}.

\section{ Related Works}
\label{related works}
\begin{table*}[ht!]

\centering
\caption{ Comparison of Related Literature on SD-SG Network Security: \\
\checkmark designates the topic is covered, $\ast$ designates the topic is partially covered, and \\
--- designate the topic is not covered}

\begin{tabular}{|c|c|c|c|c|c|c|c|}

 \hline
 \label{Related_Works_Table}
 References & \specialcell{Publication\\ Year} & \specialcell{DDoS \\Attacks} & \specialcell{Controller\\ Attacks} & \specialcell{Intrusion \\Attacks} & \specialcell{Defense Techniques \\for each Cyberattack} & \specialcell{Multiple\\Attacks}& \specialcell{Emerging \\ Threats} \\
 \hline
 \cite{rehmani2019software}   & 2019  &--- & $\ast$ & $\ast$& $\ast$ &--- & ---  \\ \hline
 \cite{ibdah2017security}& 2017  & $\ast$  &--- & ---&---&--- & ---\\ \hline
 \cite{akkaya2015software} & 2015 &---&$\ast$  & \checkmark &$\ast$ & --- & ---\\ \hline
 \cite{abujubbeh2019software} & 2019 &$\ast$  &--- &---& $\ast$& --- & ---\\ \hline
 \cite{demirci2018software}&   2018  & \checkmark  & ---  &$\ast$&$\ast$ & --- & ---\\ \hline
 \cite{kim2015trends}& 2015 &---&----&$\ast$& $\ast$& --- & ---\\ \hline
 
 \textbf{This Survey} & 2023  & \checkmark & \checkmark &\checkmark &\checkmark & \checkmark & \checkmark\\  \hline
 
 \hline
\end{tabular}
\end{table*}



 
 %

This survey is novel from other surveys because it is the first of its kind to provide an in-depth examination of network security threats and defense methods for SD-SG. Related literature have only briefly or completed forgone covering these topics in depth. Table~\ref{Related_Works_Table} and  summarizes the comparison of this work with other related literature and surveys.
Table~\ref{Acrynonyms_Definitions} presents a list of acronyms and their definitions used throughout the paper.
Table~\ref{Related_Survey_Articles_SG_and SD-SG} summarizes related literature for both SG and SD-SG security. This paper concentrates primarily on SD-SG network security and persistent threats, therefore related work that focuses on SG security in general, as seen in the table, is outside its scope. Furthermore, in this paper, we focus mainly on related literature published within the last five years of the paper's publication date to ensure that we survey the most recent, novel literature and solutions. This section provides a review of related literature and surveys of SD-SG and highlights their missing components which are found in this survey.

\begin{table}[ht]

\centering
\caption{ List of Acronyms and Definitions}

\begin{tabular}{|c|c|}

 \hline
 \label{Acrynonyms_Definitions}
 \textbf{Acronyms} & \textbf{Definitions} \\
 \hline
 SG & Smart Grid \\ \hline
 SD-SG & Software-Defined Smart Grid  \\ \hline
 SDN   & Software-Defined Networking   \\ \hline
 DDoS & Distributed Denial-of-Service \\ \hline
 LDoS & Low Rate Denial of Service \\ \hline
 ICT & Information and Communication Technologies \\ \hline
 SDWSNs & Software-Defined Wireless Sensor Networks  \\ \hline
 HANs & Home Area Networks \\ \hline
 NANs & Neighborhood Area Networks \\ \hline
 WANs & Wide Area Networks \\ \hline
 APIs & Application Programming Interface \\ \hline
 ForCES & Forwarding And Control Element Separation \\ \hline
 PCEP & Path Computation Element Communication Protocol \\ \hline
 NetConf & Network Configuration Protocol \\ \hline
 I2RS & Interface to Routing System \\ \hline
 FML & Flow-Based Management Language \\ \hline
 RESTful & Representational State Transfer \\ \hline
 ALTO & Application Layer Traffic Optimization \\ \hline
 NVP & Nicira Network Virtualization Platform  \\ \hline
QoS & Quality of Service \\ \hline
OVSDB &  Open vSwitch Database Management \\ \hline
BC & Blockchain \\ \hline
POF  & Protocol Oblivious Forwarding \\ \hline
P2P & Peer-to-Peer Communication \\ \hline
RNNs & Deep Recurrent Neural Networks \\ \hline
BiLSTM & Bidirectional Long Short RNN \\ \hline
SCADA & Supervisory Control and Data Acquisition \\ \hline
MTD & Moving Target Defense \\ \hline
IDS & Intrusion Detection Systems \\ \hline
HIDS & Host IDS \\ \hline
SIDS & Signature-Based IDS \\ \hline
AIDS & Anomaly-Based IDS \\ \hline
ML & Machine Learning \\ \hline
SD-CPC & Software-Defined Controller Placement Camouflage \\ \hline
VSF & Virtual Security Functions \\ \hline
RED &  Random Early Detection \\ \hline
TCP & Transmission Control Protocol \\ \hline
AQM & Active Queue Management \\ \hline
C\&C & Commmand
and Control Channel \\ \hline
DNS &  Domain Name System \\ \hline
DDNS & Dynamic DNS \\ \hline
WSN & wireless sensor networks \\ \hline
MANETs & mobile ad hoc networks \\ \hline
SDDCs & software-defined data centers \\ \hline
CECD-AS & Cross-Layer Ensemble CorrDet with Adaptive Statistics \\ \hline
FDI & False Data Injection \\ \hline
TCP-SYN & Transmission Control Protocol-Synchronize \\ \hline
TSA & Time Synchronization Attack \\ \hline
MITM & Man-in-the-Middle \\ \hline

\end{tabular}
\end{table}

Rehmani et al. \cite{rehmani2019software} present a comprehensive literature review of SDN in SG communications. The authors discuss the challenges posed by SGs and how SDNs can assist in overcoming them. This paper examines the benefits of SDN-based communication for SGs, including enhanced network performance, resource management, and security. The authors provide a comprehensive analysis of existing SG communication solutions based on SDN, including their architectures, protocols, applications, and security. Additionally, the paper identifies unresolved research issues and challenges that must be addressed to facilitate more efficient use of SDN-based communication in SGs. It provides insights into technologies and identifies areas for future research and development. However, the paper's security section mentioned within the paper only provides a brief overview of the security of SD-SG, and the security papers covered are now outdated. Furthermore, it does not provide an in-depth discussion of future research efforts for SD-SG security threats as mentioned in this paper.

\begin{table*}
\centering

\caption{Comparison of Survey Articles of Smart Grid and Software Defined Smart Grid Security}
\begin{tabular}{|p{4.5cm}|p{4.5cm}|p{4.5cm}|}

\hline
\label{Related_Survey_Articles_SG_and SD-SG}
Main Domain & Sub-Topic: Cyberattack & References \\
\hline
\multirow{17}{*}{Smart Grid Security} & 
DDoS/DoS/Physical-DoS (PDoS)  & \cite{hahn2011cyber,pedramnia2018survey,lee2020study, feher2020smart}\\\cline{2-3}
&Spoofing, Sniffing, and Message Relay & \cite{wang2011analysis, ali2013randomizing, rajkumar2020cyber, mohan2020distributed}\\\cline{2-3}
&MITM,Eavesdropping, and Homograph & \cite{baig2013analysis, fritz2019simulation, wlazlo2021man, khan2022elliptic, el2018cyber}\\\cline{2-3}
&Meter Manipulation and Theft & \cite{farokhi2020review, kim2019detection, singh2021blockchain, han2016fnfd}\\\cline{2-3}
&FDI & \cite{musleh2019survey, duan2016resilient, chung2017local, jiang2017real}\\\cline{2-3}
&Impersonation, Session Key Exposure, and TSA & \cite{srivastava2021emerging, wu2019anonymous, chaudhry2022sg, ebrahimabadi2021hardware, chen2022anonymous, shereen2019model}\\\cline{2-3}

&TCP-SYN Flooding & \cite{bogdanoski2013analysis, holik2022threat,ansilla2015data, kwon2015behavior}\\\cline{2-3}
&Jamming & \cite{gunduz2020cyber, gai2017spoofing, ma2015multiact, lu2014camouflage, zhang2022jamming, xu2021cooperative}\\\cline{2-3}
&RAM Exhaustation/CPU Overload & \cite{  chen2012smart, sun2020intrusion, fadlullah2011early}\\\cline{2-3}
&Brute Force & \cite{cairns2013flexible,nge2019real, nicanfar2011smart,sha2016secure}\\\cline{2-3}
&Message Replay, Covert & \cite{tran2013detection, farraj2017distributed,pavithra2021prevention,li2013efficient, tanveer2021robust, ahmed2018feature}\\\cline{2-3}

&Sybil & \cite{wei2011protecting, najafabadi2013sybil, sriranjani2023received, kumari2020performance}\\\cline{2-3}
&Multi-Attack &\cite{feher2020side, ali2014two,gayathri2019multi, abou2022ensemble, sakhnini2021physical
}\\\cline{2-3}

\hline
\multirow{2}{*}{Software-Defined Smart Grid Security}
& DDoS/DoS & \cite{xiong2021distributed, polat2022novel, nagaraj2021glass, abdelkhalek2022moving, mahmood2021s } \\ \cline{2-3}
&Controller & \cite{sivaraman2020game, samir2021sd, lin2019security, azab2022mystify }\\\cline{2-3}
&Intrusion & \cite{info10030106, 9664737, jung2019anomaly, 9843645, 10.1007/978-3-030-02931-9_7, Peng, 10017381, 8746112  } \\\cline{2-3}
&Multi-Attack & \cite{starke2022cross,agnew2022implementation,aljohanicross,nagaraj2020ensemble,trevizan2019data,ruben2020hybrid} \\\cline{2-3}
\hline
\end{tabular}
\end{table*}

Ibdah et al. \cite{ibdah2017security} discuss the vulnerabilities of using SDN in SG systems. The authors start by outlining the architecture of SDN-based SG systems as well as the security risks associated with this technology. The paper then proposes a five-component security framework for SDN-based SG systems: secure communication, secure data storage, secure computation, secure authentication, and secure access control. The authors go into detail about each of these components and how they can be used in an SDN-based SG system. The authors use the Mininet network emulator to simulate the effectiveness of their proposed security framework. The simulation shows that the proposed framework is effective at mitigating various types of security attacks, such as DoS attacks, data tampering, and unauthorized access. Although the authors provide discussion of security of SD-SG systems, the discussion is very brief and focuses on DDoS attacks and proposal of their framework. The researchers do not provide a comprehensive review of SD-SG security like this survey.

Akkaya et al. \cite{akkaya2015software} investigated the use of software-defined networking (SDN) in wireless local networks (WLANs) for SG applications. The authors begin by outlining the difficulties that traditional WLANs face when supporting SG applications, such as scalability, reliability, and security. The paper proposes an SDN-based architecture for WLANs in SG applications, which includes a centralized control plane capable of dynamically reconfiguring the network to meet changing needs. The authors detail the proposed architecture, including the various components and their functions and use the NS-3 network simulator \cite{riley2010ns} to simulate the effectiveness of the proposed SDN-based architecture. The simulations show that the proposed architecture outperforms traditional WLANs in terms of network performance, resource utilization, and security. The researchers investigate the relevant security aspects for each deployment scenario as they progress. However, the researchers only provide discussion of anomaly/intrusion attack detection and only provide breif discussion of controller attacks. The paper does not provide discussion of DDoS attacks nor defense techniques for the attacks mention in this paper.

Abujubbeh et al. \cite{abujubbeh2019software} discusses the application of software-defined wireless sensor networks (SDWSNs) in SGs. The authors begin by discussing the advantages of SDWSNs over traditional wireless sensor networks, such as increased flexibility, scalability, and adaptability. The paper proposes a  SDN architecture for SDWSNs in SGs to provide a centralized control plane for network management. The authors describe the proposed architecture's various components, such as sensor nodes, SDN controllers, and network infrastructure. In addition, the paper discusses the various challenges that SDWSNs face in SGs, such as security, energy efficiency, and network reliability. The authors emphasize the importance of overcoming these challenges in order to effectively use SDWSNs in SGs. In constrast to this survey, this paper only provides brief discussion of DDoS attacks and their defense techniques and negates other network security threats mentioned in this survey. Furthermore, this paper concentrates on SDWSNs instead of general SD-SG.

Demirci et al. \cite{demirci2018software} suggests implementing software-defined networking (SDN) to improve the security of SG systems. The authors discuss the challenges encountered by conventional SG systems, such as the lack of network traffic visibility and control, which makes it difficult to detect and respond to security threats. The paper proposes an SDN-based security framework for SG systems that consists of three elements: network visibility, policy enforcement, and threat detection and response. The authors provide a thorough description of each of these components and discuss how they can be implemented using SDN. To evaluate the efficacy of the proposed framework, the authors emulate their networking using the Mininet \cite{de2014using}. The simulation demonstrates that the proposed framework can enhance network visibility, policy enforcement, and threat detection and response in comparison to conventional SG systems. However, this paper focuses on proposing their framework instead of a comprehensive review of network security of SD-SG. The authors discuss DDoS attacks but only briefly address anomaly/intrusion attacks and defense techniques for each attack in contrast this survey which a more comprehensive review of network security threats. Furthermore, they provide no mention of controller attacks in current literature.

Kim et al. \cite{kim2015trends} discuss the evolution of SG infrastructure as well as the potential for SDN to enable advanced SG capabilities. The authors begin by outlining SG infrastructure and the role of information and communication technologies (ICT) in enabling advanced functionalities like demand response, distribution automation, and advanced metering. The paper provides discussion about the difficulties that traditional SG infrastructures face, such as a lack of interoperability and flexibility. The paper proposes using SDN to address the challenges that traditional SG infrastructures face while also enabling advanced functionalities. The authors describe the different components of an SDN-enabled SG architecture, such as the SDN controller, network infrastructure, and applications. The authors review the potential advantages of an SDN-enabled SG infrastructure, such as increased reliability, security, and energy efficiency. They highlight a number of use cases in which SDN can enable advanced functionality such as dynamic load balancing and network slicing. However, this paper avoids discussing DDoS and controller attacks and only briefly discusses anonymity/intrusion attacks and defense techniques, all topics covered within this survey.


\section{Background}
\label{Background}
\subsection{What are software-defined smart grids?}
\label{What are software defined smart grids?}
As aforementioned, SD-SG are a form of SG that utilizes SDN technology to better manage bus communications, network topology organization, security, and grid network visibility and control. Furthermore, SD-SG utilizes advanced data analytic with SDN and real-time monitoring to control the grid's communication layer. An high-level overview of this integration and SDN within a SG can be seen in Figure~\ref{fig:SDN_based_smart_grid_arch}. Each layer as it pertains to SD-SG can be defined as follows:
\begin{itemize}
    \item \textbf{\emph{Infrastructure/Data Layer:}} This layer is predominantly concerned with the flow of data from and to SG entities such as energy sources, servers, power transmission lines, and private/commercial consumers. This data is transmitted to programmable SDN-based switches and routers for routing to the intended location. Control layer routing information is implemented in this layer.
    \item \textbf{\emph{Control Layer:}} This SD-SG layer is primarily composed of two components: the SDN controller(s) and the advanced distribution management system. (ADMS). ADMS consists of supplementary control and data acquisition (SCADA), distributed energy resource management (DERMS), and distribution management system (DMS) in order to monitor the smart grid system. This layer receives and transmits data to the application layer.
    \item \textbf{\emph{Application Layer:}} This is the highest layer and it predominantly receives input in the form of data from the lower two layers in order to ensure that the system is operating in accordance with the established policies. To ensure the safety and stability of the SD-SG, it executes real-time monitoring, statistical analysis, traffic engineering, and security monitoring. 
\end{itemize}
More detailed informaiton of the SDN layers and inter-communication can be found in section~\ref{What is SDN?}. 

\subsection{What is software defined networking?} \label{sec_SDN}
\label{What is SDN?}


SDN has taken shape in recent years and was used to represent the ideas and techniques used for Openflow at Stanford University \cite{mckeown2008openflow, nisar2019survey}. SDN simply refers to the network architecture whereas the data plane and control plane are decoupled from one another. SDN has the following characteristics \cite{maleh2022comprehensive}:
\begin{itemize}
    \item Control plane and data plane are separated from one another.
    \item Forwarding decisions are based on flow and not destination. In SDN, a flow is defined as a series of packets between the source and destination that follow a set of instructions. The forwarding devices have the flows installed into their flow tables by controllers.
    \item The controller acts as the central logic and external entity. Its job is to direct the traffic through the network and maintain the status of the network.
    \item The network has the ability to be programmed through software applications running on top of the SDN controller.
    \item Application programming interface (APIs) are used to pass information between planes of the SDN infrastructure.
\end{itemize}

Although SDN was first launched publicly, the concept of SDN has been evolving since 1996, with the objective of enabling user-controlled management of forwarding in network nodes\cite{zhang2022performance}. There have been numerous stepping stones before the eventual creation of SDN technology as we know it today. The General Switch Management protocol in 1996, the Tempest (framework for programmable networks) in 1998, and  most recently,  Ethane and OpenFlow, in 2007 and 2008 respectively, were necessary to make modern SDN a reality. 
\begin{figure}[h]
\centering
\includegraphics[width=\linewidth,trim = 0cm 0mm 0cm 0cm]{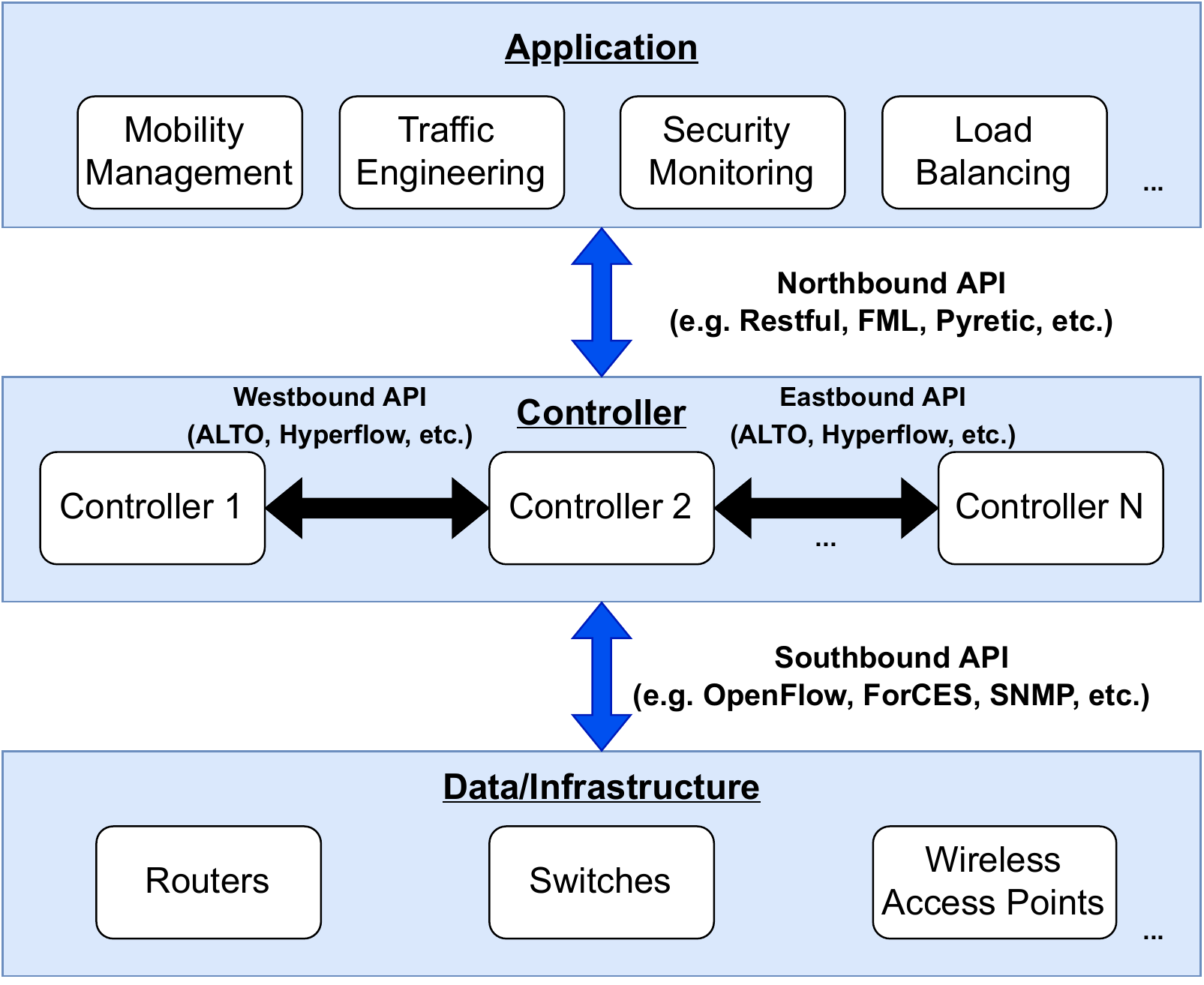}
\caption{General SDN Architecture}
\label{fig:SDN_arch}
\end{figure}
 Figure~\ref{fig:SDN_arch} represents an overview example of a modern functional SDN architecture. Modern SDN features include the following: 
 \begin{itemize}
     \item Centralized controller
     \item Separation of the control and data planes
     \item Open interface between controller and data plane
 \end{itemize}
 As shown in figure~\ref{fig:SDN_arch}, the infrastructure layer is composed of routers, switches, and access points. This represents the physical network equipment in the network and this layer forms the data plane. The controller communicates with the data plane using southbound  programming interfaces (API) such as OpenFlow~\cite{kreutz2014software}, ForCES~\cite{haleplidis2015network}, PCEP~\cite{vasseur2009path}, NetConf~\cite{enns2006netconf}, or I2RS~\cite{hares2013software}. If there is more than one controller present, these controllers communicate with each other using Westbound and Eastbound API such as ALTO~\cite{zhou2014rest} or Hyperflow~\cite{tootoonchian2010hyperflow}. The topmost layer is the application plane. In this layer, the network operator is able to utilize functional applications for various tasks such as energy efficiency, access control, mobility management, and/or security management. The application layer communicates with the control layer using the Northbound APIs such as FML~\cite{hinrichs2009practical}, Procera~\cite{voellmy2012procera}, Frenetic~\cite{foster2011frenetic}, and RESTful~\cite{zhou2014rest}. Depending on the desired results, the network operator can send the necessary changes to the control layer using these APIs so that the controller can make necessary changes in the infrastructure layer. 
 \begin{figure}[h]
\centering
\includegraphics[width=\linewidth]{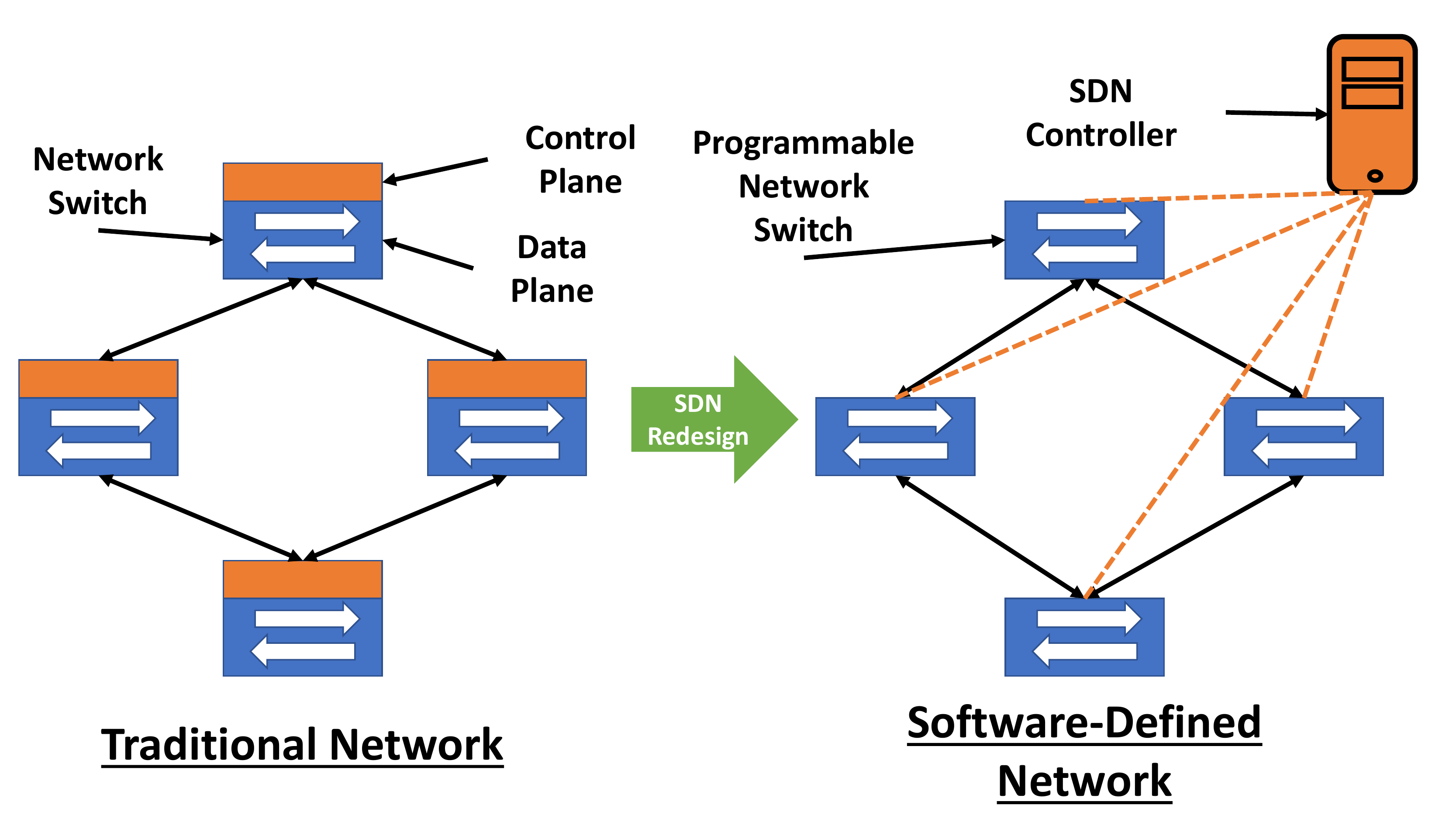}
\caption{SDN vs. Traditional Approach for Networking }
\label{fig:New_vs_Old_SDN}
\end{figure}

 \subsection{SDN versus Traditional Networking}

 There are unique differences between traditional approaches to networking compared to SDN. As seen in \cref{fig:New_vs_Old_SDN} and discussed in\cite{haji2021comparison}, the control plane and data plane are connected in the traditional network architecture while in SDN the control and data plane are in a different stack. In SDN, the control plane bears the responsibility to configure the nodes and defined the paths for the data flows. Once the paths are determined, the controller then gives this information to the data plane. However, in traditional networks, the flow management, or flow policy, is defined by the forwarding devices. The only way to change this policy is by physically reconfiguring the devices themselves, which can be restrictive. Because of these hurdles, network management policies in traditional networks are not as dynamic and are difficult to scale. SDN offers the ability to quickly change data flows, enabling the network operator to meet the changing traffic demands. 
  

Important terms to note in SDN are:
\begin{itemize}
    \item \textbf{\emph{Forwarding devices:}} These are devices that can be hardware or software-based and do the routine tasks for incoming packets such as forward to specific ports, drop, or forward to the controller. They are controlled by Southbound API as shown in \cref{fig:SDN_arch}.
    
    \item \textbf{\emph{Data plane:}} Connected forwarding devices in the network represent the Data plane.
    

  \item\textbf{\emph{Southbound API:}} These are the instructions that are sent from the controller (control plane)  to the switches. The most common southbound API is OpenFlow.
  
  \item\textbf{\emph{East/West API:}} If more than one controller is used, the controllers communicate with each other using East/West API. These API are used by controllers to update each other on conditions and statuses of devices over which they preside. This allows the controllers to maintain a global view of the network regardless of the topologies over which the controllers maintain.  
  
  \item \textbf{\emph{North Bound API:}} This is the communication between the SDN controller and the application layer. 
    
  \item \textbf{\emph{Application plane:}} This plane is a set of applications that set the policies of the network. An application can be routing, firewalls, load balancers, etc \cite{al2020comprehensive}. By using northbound API, the application layer gives the controller instructions on what behaviors the network should follow.
\end{itemize}

\subsection{Data/Infrastructure Plane}
SDN and traditional networks are alike in the sense that they both contain networking equipment, such as routers and switches. The difference is that in SDN those devices no longer make autonomous decisions of forwarding like in traditional networks. The devices act as simply forwarding devices and receive instructions on how to route traffic from the SDN controller. This moves the intelligence of the network from the data plane devices to the controller. Furthermore, these networks contain communication protocols and APIs (i.e. OpenFlow) to ensure the right configurations of the network. This allows the SDN controller immense control and the ability to make adaptive changes to the network in response to the demands of the network operator. 

A data plane device is a hardware or software device that forwards packets through the network. An OpenFlow device utilizes flow tables to direct traffic. The data in these flow tables are a matching rule, actions to be executed on packets that match these rules, and counters to keep statistics of how many packets match these rules. These flow tables define what will happen to packets received to these forwarding devices. When a new packet arrives at the forwarding device, the device checks its flow table to see if there are established instructions on how to properly handle the packet. If there are no flow, or instructions, in the flow table, the forwarding device will send the packet to the controller using the OpenFlow protocol for it to determine what should be done with it. The controller will inspect the packet and determine the appropriate forwarding action.

\subsection{Control Plane/Controller}

The controller layer is crucial to any SDN network and provides complete control over the network. Controllers allow for the user to control the flow of packets throughout the network by creating or deactivating links between nodes. The controllers gather information from the equipment on the health of the network. 
It allows for ease of network management by providing a central point of control for the network. The controller allows the network management with high-level abstraction and APIs to developers. The controller can provide information such as network topology, flow statistics, and device discovery. While using an SDN controller, a developer does not need to be concerned with low-level details of data distribution amongst forwarding devices. 

Many projects have been launched to construct controllers due to the usability and importance of controllers for the SDN control plane. SDN controllers can be separated into two categories centralized vs decentralized \cite{kreutz2014software}\cite{ahmad2021scalability}.
\subsubsection{Centralized} Centralized controllers were the first controllers to arise with the first version being NOX\cite{gude2008nox}. NOX was released in 2008 and was the first OpenFlow controller. Having a centralized controller provides numerous benefits compared to traditional networks. In traditional networks, each forwarding device has very little information about the global state of the network. SDN controllers, on the other hand, have information about network topology, traffic flows, and switch load \cite{ahmad2021scalability}. The controller sees the entire network and is able to quickly deploy to manage or change any area of it. It can also route optimal paths through the network since it has a global view of the routing paths. In traditional networks, devices must be added to do certain tasks such as firewalls, intrusion detection, load balancing, etc., but in SDN, all of these tasks can be done by implementing an application  in the application layer\cite{ahmad2021scalability}\cite{chica2020security}. These effectively remove the need for these devices and reduce the cost and ease of implementation needed to run the network under desired constraints. In addition, a central controller makes it easy for a developer to apply an application to the entire network through one singular controller instead of having to communicate to multiple controllers in a distributed architecture. 

Although there are multiple advantages of using a centralized controller for an SDN topology, there are certain limitations and disadvantages as well. For a large wide area network (WAN),  using a single SDN controller may not be sufficient to properly maintain the network. The network will experience high control latency and bandwidth issues since one controller has to respond to multiple nodes over a large area \cite{ahmad2021scalability}. Furthermore, as the number of nodes increases in an SDN topology, these nodes will all need to communicate regularly with the controller. This increased number of dynamic communication, in short, the substantial bursts can overload a single controller. Therefore, to handle the load of a WAN, it is beneficial to design a distributed approach and add more controllers to the network. 

\begin{figure}[h]
\centering
\includegraphics[width=\linewidth]{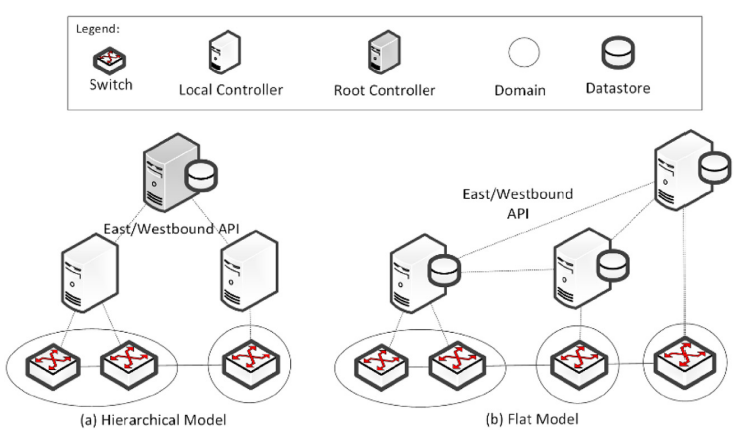}
\caption{Distributed Topologies \cite{haggag2021network}}
\label{Distributed Topologies_APIs}
\end{figure}

\subsubsection{Distributed} \label{Distributed SDN} In recent years, there has been a surge of interest in distributed control plane topologies \cite{kreutz2014software}\cite{bannour2017distributed}. They have addressed several of the restrictions that centralized control planes have, such as scalability, dependability performance, and single point of failure \cite{ahmad2021scalability}. The distributed controller SDN structure enables the control plane to respond to various locations in the network in an efficient manner. It enables more dynamic network control and maintenance. Furthermore, the distributed controller's scalability and flexibility make it perfect for WAN.

However, decentralized topologies are not without their challenges. Distributed controllers use load balance and share metrics about the domains they are connected to with each other. This East/West API communication as seen in \cref{fig:SDN_arch} is crucial for having a logically center, but physically distributed topology. Two examples of distributed topologies can be seen in Figure~\ref{Distributed Topologies_APIs}. The two distributed topologies shown are hierarchical and flat models. In the hierarchical model, the root controller has global knowledge of the network and is in charge of the local controllers. The local controllers do not communicate with each other in the hierarchical model. On the other hand, the flat model does not contain any root controller and relies heavily on East/West API communication to share with each other the global state of the network. In both networks, there is a degree of difficulty that is increased when there are multiple controllers that need to communicate with one another. In the hierarchical model, the root controllers provide  a single point of failure. In the flat model, challenges arise with the need for regular communication needed between controllers to update the global state of the network\cite{chahlaoui2020taxonomy, ahmad2021scalability}.


East and westbound are essential in distributed SDN topologies. They represent the communication between controllers as they update each other on the state of the network. Figure~\ref{Distributed Topologies_APIs} represents this communication. Import/export data across controllers, methods for data integrity models, and monitoring/notification capabilities (e.g., check if a controller is up or notifying a take over on a set of forwarding devices) are all features of these interfaces\cite{alsaeedi2019toward}. East and westbound APIs are crucial to provide common compatibility and interoperations between controllers in a system.

\subsection{Southbound API}
Southbound API connects the data plane, or infrastructure, with the control plane. They provide a bridge of communication between the two layers. There are many southbound APIs but OpenFlow is the most widely accepted one \cite{chica2020security}. It provides a common standard and procedure for OpenFlow-enabled forwarding devices and controllers. OpenFlow supports three data streams according to \cite{kreutz2014software}:
\begin{itemize}
    \item Event-based messages are sent to forwarding devices to the controller when a link or port change happens.
    \item When a link or port change occurs, event-based messages are transmitted to the controller via forwarding devices.
    \item When a forwarding device is unsure what to do with a packet, it transmits it to the controller for additional processing. 
\end{itemize}

Although OpenFlow is the most common, there are other available southbound APIs such as ForCES, OVSDB, and POF. ForCES keeps the architecture of the network the same as traditional networks but still separates the control and data planes. However, it does not need  a centralized controller and the control plane is updated using third-party firmware on the devices.
Open vSwitch Database Management (OVSDB) has more advanced options for managing the network compared to OpenFlow. It allows the user to create multiple virtual switches, set the quality of service (QoS) standards for the network, attach interfaces to the switches, configures tunnel interfaces on OpenFlow data paths, manages queues, and collect statistics \cite{kreutz2014software}. Essentially, OVSDB is seen as a more advanced, complementary option to OpenFlow.
Protocol Oblivious Forwarding (POF) was created as a direct competitor to OpenFlow inorder to improve the SDN forwarding plane. In OpenFlow, the switches have to extract certain identifying bits to match the packet with the correct flow table entry. However, this can be computationally expensive and increase complexity. POF tries to circumvent this by making the forwarding devices protocol oblivious. The forwarding devices do only processing and forwarding. Packet parsing is left to the controller to do.

\subsection{Northbound API}

Northbound APIs are  used for the control layer to communicate with the application layer. There is a variety of different Northbound bound APIs available such as ad-hoc APIs, RESTful APIs, NVP\cite{haggag2021network, haggag2019network, humernbrum2014using}. The northbound interface between the controller and application layer is mostly software-based unliked the southbound API which is mostly hardware-based. 

\subsection{Application Layer}

The application layer is implemented on top of the control layer. It is responsible for giving the SDN controller commands to implement to the data plane layer. Since SDN can be applied to a wide range of topologies and environments, it can be used for a wide range of applications, including routing applications, load balancing, security enforcement, fail-over, reliability functionalities in the data plane, end-to-end quality of service (QoS), network virtualization, and mobility management in wireless networks \cite{kreutz2014software}. These applications can check that SDN topologies fulfill the developer's specifications. Furthermore, these applications have made SDN appealing for a wide range of network requirements.

\subsection{SD-SG Cyberattack Threats}
\label{sdnsecurity}

\begin{figure*}[ht]
\centering
\includegraphics[width=1.8\columnwidth, trim = 1cm 10cm 0cm 1cm]{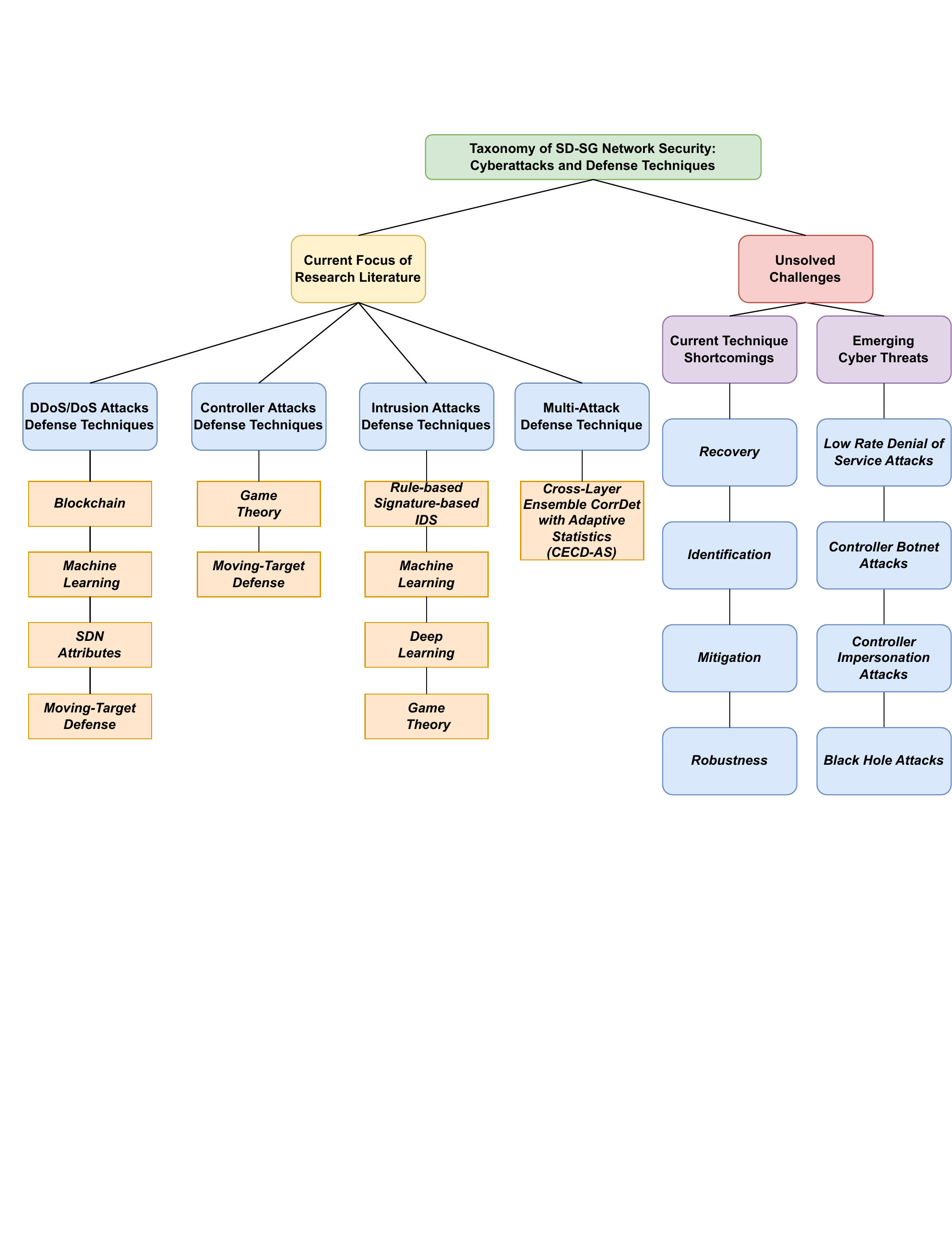}
\caption{Taxonomy of SD-SG Network Security: Cyberattacks and Defense Techniques}
\label{fig:taxonomy_SD-SG_attacks}
\end{figure*}

Research in the field of SD-SG security has examined a variety of different attacks. From a thorough investigation of the literature of SDN based security protocols, the main attack types that are present in this literature fall into the following categories as shown in Figure~\ref{fig:taxonomy_SD-SG_attacks}:
\begin{itemize}
    \item \textbf{\emph{Distributed Denial-of-Service (DDoS):}} DDoS/DoS attacks are types of cyber-attacks in which the attacker attempts to consume all of the server's resources so that the server is unable to respond to legitimate requests by launching a multi-node attack on a victim\cite{ahmed2019detection}.
    \item \textbf{\emph{Controller Attacks:}} SD-SG controllers in SDN networks are vulnerable to a variety of attacks, including DoS, hijacking, unauthorized access, etc. \cite{santos2020machine, maleh2022comprehensive}. These attacks seek to exploit SDN's centralization of control and disrupt the network by attacking the controller. In singular controller frameworks, the controller can serve as the network's single point of failure.
    \item \textbf{\emph{Intrusion Attacks:}} Anomaly/Intrusion attacks are cyber-attacks where the attacker gains unauthorized access \cite{sultana2019survey}. After gaining illicit access to the system, the attackers may launch other attack such as DoS, eavesdropping, controller attacks, packet manipulation, etc. 
    \item \textbf{\emph{Multiple Attacks:}} Research has been developed to detect multiple attacks launched against the SD-SG such as the aforementioend attacks as well as other attack types such as false data injection (FDI) and man-in-the-middle (MITM) attacks by using a cross-layered approach. According to Allen et al. \cite{starke2022cross}, FDI cyberattacks attempt to deceive the control algorithm by injecting forged data into the network, whereas an MITM attack attempts to gain access to the communications between two nodes in a system by impersonating a node. During an MITM attack, the attacker may attempt to read the contents of communications or manipulate packets.

\end{itemize}


\section{Network Security Challenges of Software-Defined Smart Grids}
\label{Network Security Challenges of Software-Defined Smart Grids}

This section contains a survey that details current network security challenges in the form of cyberattacks on SD-SG and their countermeasures. We survey and analyze DoS/DDoS attacks, controller attacks, anomaly/intrusion attacks, and multiple attack defense techniques in current literature. An high level overview of the taxonomy of these attacks and their defenses surveyed from literature can be seen in Figure~\ref{fig:taxonomy_SD-SG_attacks}.

\subsection{DDoS/DoS Attack Defense}
\label{Distributed Denial of Service Attack Defens}

DDoS attacks can have devastating consequences for SD-SG \cite{agnew2022implementation, starke2022cross}. These attacks can be carried out by malicious nodes that have been hijacked by attackers or by nodes that have gained unauthorized access to the framework. DDoS attacks have grown in popularity among cyberattackers attacking SDN frameworks \cite{dantas2020taxonomy} due to their ease of use. As a result, SD-SG network operators and researchers have had to consider DDoS attack mitigation strategies. The current literature and efforts for DDoS attack detection and mitigation for SD-SG are presented in this section and categorized based on the following:
\begin{itemize}
    \item \textbf{\emph{Blockchain:}} Blockchain (BC) is a potential solution for enforcing a verification system at every edge of SD-SG's SDN networks and addressing trust-management issues in order to make the system resilient to various attacks and guarantee block validation using encryption or consensus mechanisms \cite{rahman2022integration,mollah2020blockchain }. BC offers a peer-to-peer (P2P) communication system that keeps track of every connection-related record for each network member in an easily accessible and properly kept database.
    \item \textbf{\emph{Machine Learning:}} Machine learning developed from a collection of powerful AI techniques and has been widely used in data mining, allowing the system to learn useful structural patterns and models from training data, making it useful for networks security \cite{xie2018survey}.
    \item \textbf{\emph{Moving-Target Defense (MTD):}} MTD increases ambiguity and complexity for system attackers, decreases their chances of identifying targets (e.g., vulnerable system components), and raises the cost of attacks and scans \cite{cho2020toward}. (e.g., reconnaissance attacks). 
    \item \textbf{\emph{SDN Attributes:}} SDN permits more flexibility, control, and security as was stated in section~\ref{Background} of this survey. Researchers have made use of these characteristics to develop security models for SD-SG that make use of SDN-specific characteristics such as centralized controllers, customizable data plane flows, and natural SDN network resilience. 
\end{itemize}

\subsubsection{Blockchain} 
\begin{figure*}[h]
\centering
\includegraphics[width=2\columnwidth]{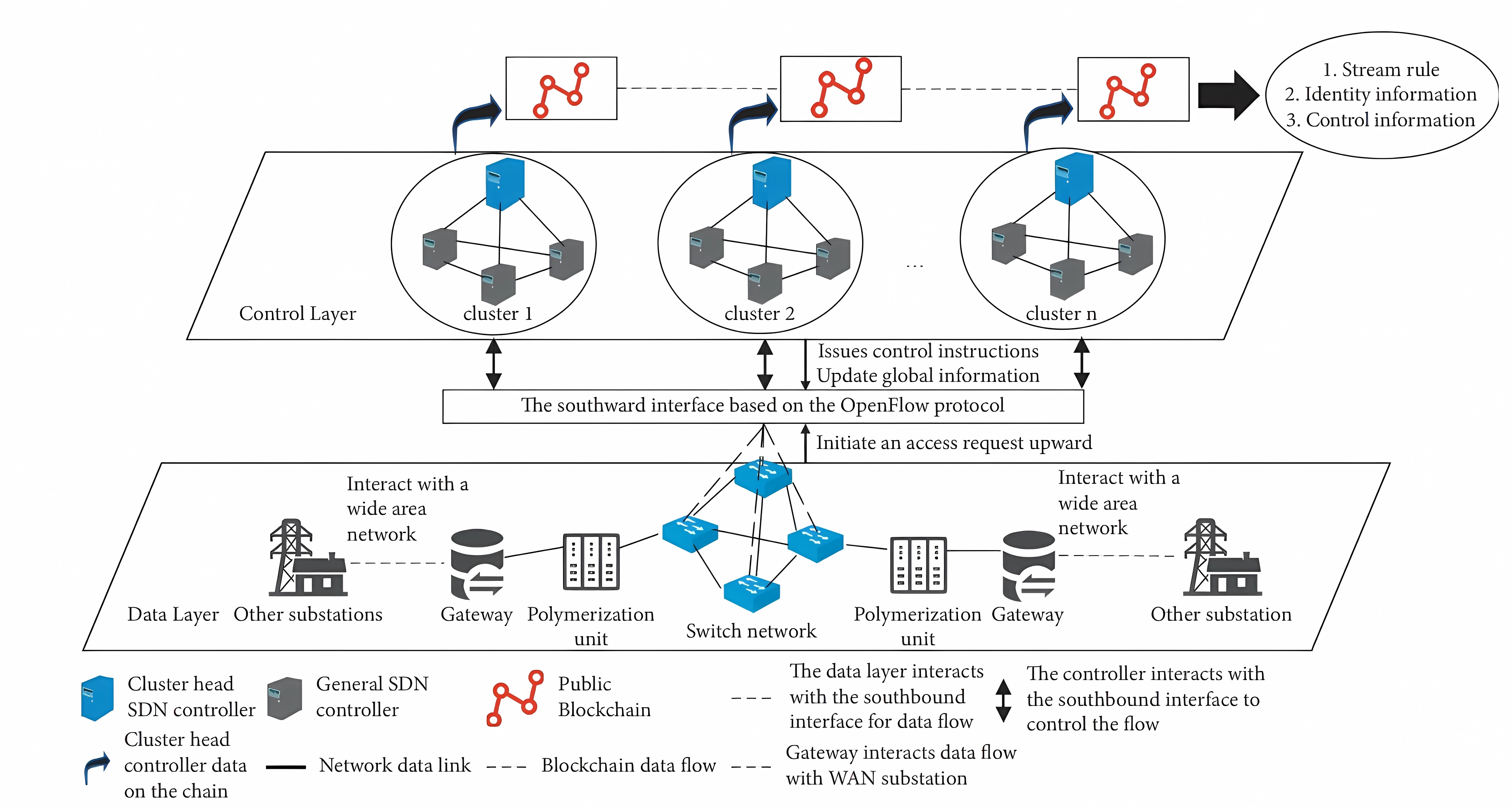}
\caption{ClusterBlock Design presented in \cite{xiong2021distributed}  }
\label{fig:ClusterBlock Design}
\end{figure*}

Xiong et al. \cite{xiong2021distributed} propose a SG distributed security architecture based on blockchain and SDN technology. Figure~\ref{fig:ClusterBlock Design} shows the suggested architecture, which makes use of blockchain technology to increase security and dependability. The proposed architecture employs a distributed consensus mechanism based on blockchain technology to ensure system integrity, prevent attacks, and provide fault tolerance and scalability, among other advantages. In addition, the authors conducted a simulation experiment to assess the performance of the proposed architecture. The proposed architecture and its results demonstrated that it has the potential to significantly improve the safety and dependability of SD-SG systems and achieve a high level of fault tolerance and scalability. The blockchain-based distributed consensus mechanism was able to protect the system from DDoS attacks and guarantee its integrity. During DDoS attack rates of 400 packets/s and 1400 packets, their framework is able to maintain bandwidth greater than $>80\%$. The study's findings indicated that incorporating blockchain technology into SG systems has the potential to improve both security and resilience when faced with DDoS attacks.

\subsubsection{Machine Learning}Polat et al. \cite{polat2022novel} proposed an approach for accurately detecting DDoS attacks in SDN-based SCADA systems using parallel recurrent neural networks (RNNs). Their proposed method utilizes a long short-term memory (LSTM) and gated recurrent units (GRU) RNN modesl to classify network traffic and accurately detect DDoS attacks that target SDN-based SCADA systems. The authors evaluated the performance of the proposed approach using a real-world SCADA network dataset. They obtained an average accuracy of 97.62\% by incorporating transfer learning which improved their model performance by around 5\%, demonstrating its effectiveness in enhancing the security of SDN-based SCADA systems against DDoS attacks.

Nagaraj et al. \cite{nagaraj2021glass} proposed GLASS, a graph learning approach for enhancing the security of SDN-based SG systems against distributed denial-of-service (DDoS) attacks. The proposed method employs graph convolutional neural networks (GCNNs) to detect DDoS attacks in real-time by learning the patterns of normal and malicious network traffic. The authors tested the proposed approach's performance on a simulated SG network and obtained a detection rate of $>97.01\%$ and maintain throughput of 84\% compared to 4\% during DDoS attack scenarios. The research shows that the proposed Glass approach improves the security of SD-SG systems against DDoS attacks.


\subsubsection{Moving Target Defense} 
\begin{figure}[h]
\centering
\includegraphics[width=\linewidth]{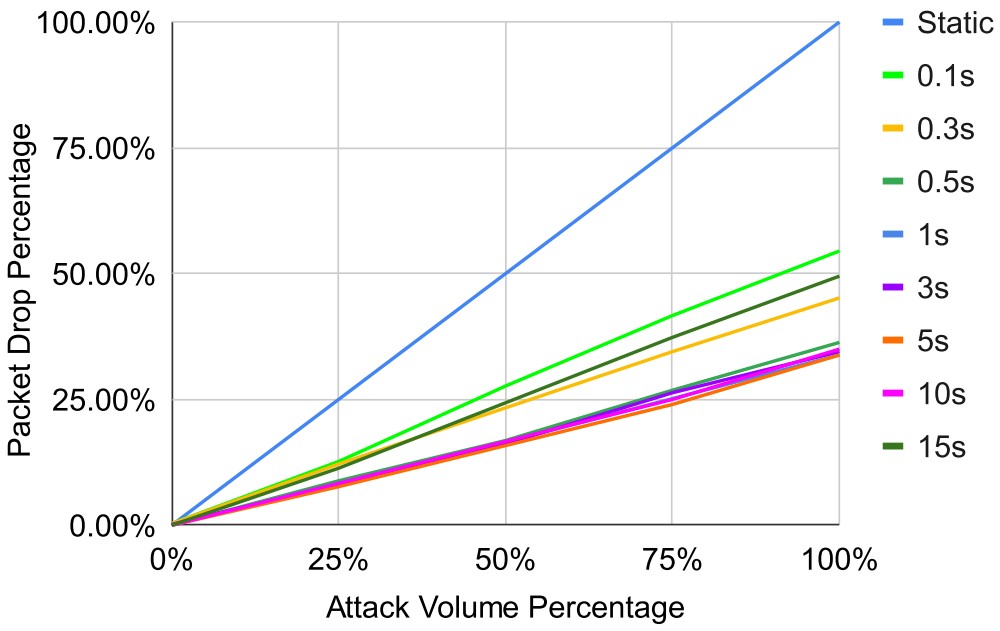}
\caption{Percentage Packet Drop vs Attack Volume Percentage for Various
MTD Switching Times as Presented in~\cite{abdelkhalek2022moving}.}
\label{fig:Packet Drop}
\end{figure}

\begin{figure}[h]
\centering
\includegraphics[width=\linewidth]{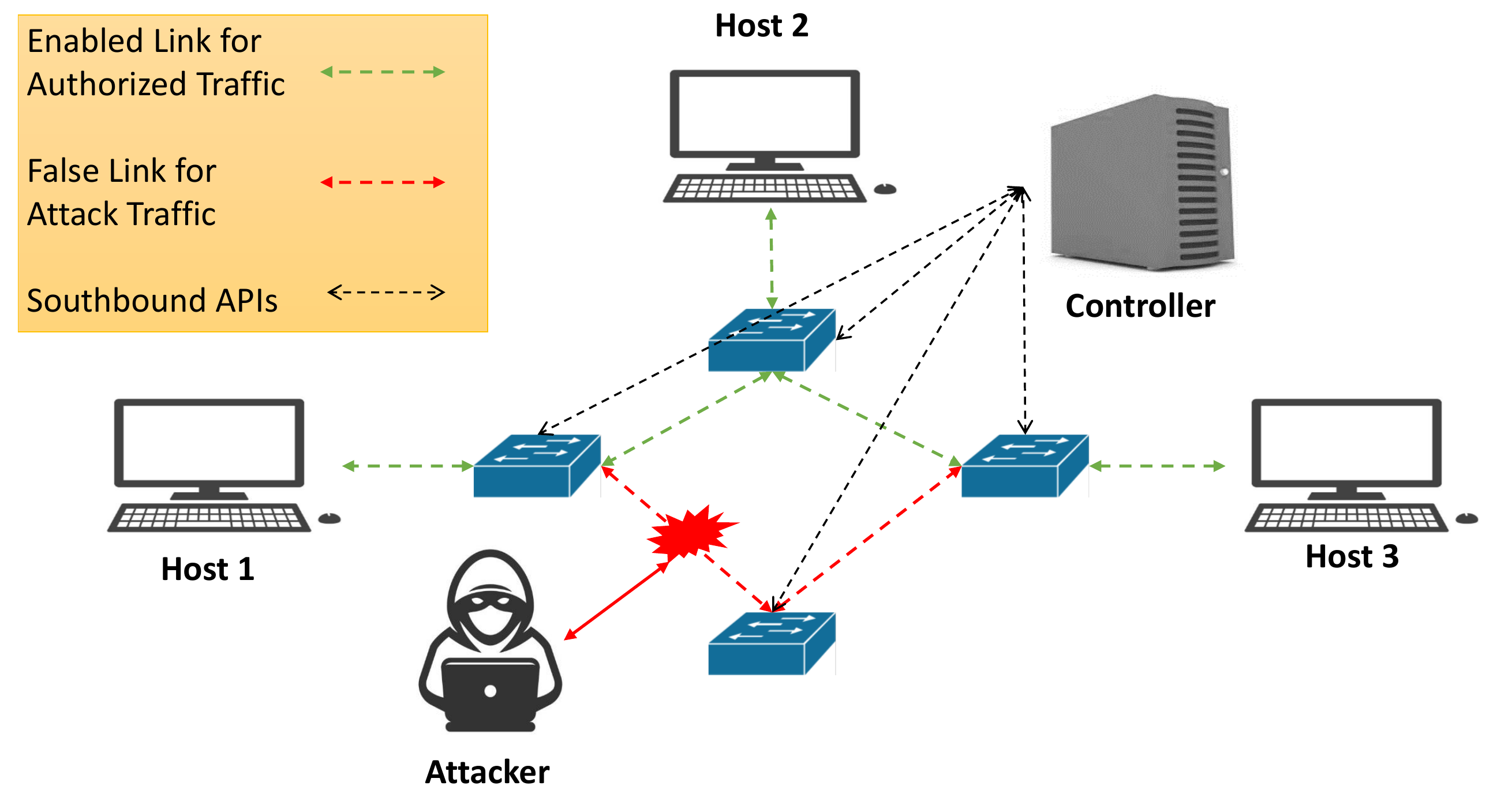}
\caption{Moving Target Defense Example Architecture}
\label{fig:CPS Model}
\end{figure}

Abdelkhalek et al. \cite{abdelkhalek2022moving} proposed a Moving Target Defense (MTD) routing mechanism to improve the security of SDN-enabled SG systems. An example of a MTD is as shown in Figure~\ref{fig:CPS Model}. Their proposed mechanism will randomize network topology by changing the paths that network flows take on in response to an attack which will make it difficult for attackers to launch successful DoS attacks. The authors evaluated the performance of the proposed mechanism using a simulated SG network demonstrate the benefits of using MTD-based routing in an SDN WAN, which has significantly reduced packet drop percentages. Researchers concluded that a link switching time of $5$ seconds was optimal for minimizing packet drops during DoS attacks as shown in Figure~\ref{fig:Packet Drop}.

\subsubsection{SDN Attributes} Mahmood et al. \cite{mahmood2021s} present S-DPs, an SDN-based DDoS protection system that can provide SG security. To protect the network, the proposed system employs a centralized controller to monitor the network and detect and classify the DDoS attacks based on  employing light-weight Tsallis entropy-based defense mechanisms, and apply countermeasures such as rate limiting and filtering. The S-DPs system was evaluated by the authors in a simulated SG environment with various attack scenarios. The results demonstrated that the system can detect and mitigate DDoS attacks with up to 100\% accuracy with a 0\% false positive rate. According to the authors, the proposed system can provide an efficient and effective approach to protecting SG from DDoS attacks.

\subsubsection{Final Thoughts for DDoS/DoS Attack Defense}
DDoS attacks pose a serious threat to SD-SG systems and are increasingly being used by cyber attackers due to their ease of use. To mitigate these attacks, several strategies have been proposed, including the use of blockchain technology, deep recurrent neural networks, access control and rate limiting, distributed software-defined networking architecture, and graph learning approaches. These strategies have yielded promising results in terms of improving SD-SG system security against DDoS attacks.

\subsection{Controller Attack Defense}
\label{controller attack defense}
Controllers are one of the most important components of any SDN framework. With access to the controller, a network operator will have complete control over network topology and traffic rules for their applications which make them an appealing target for cyberattackers for these reasons \cite{zainudin2022towards}. SD-SG network operators and designers must devise methods to keep these controllers secure so that their network is not compromised by malicious actors. This section examines SD-SG controller attack defense strategies and categorized them based on game theory and MTD. In this section, we consider MTD as described in section~\ref{Distributed Denial of Service Attack Defens}, and we define game theory as follows:
\begin{itemize}
    \item \textbf{\emph{Game Theory:}} Designers use models to analyze data and experiences. According to game theory, multi-decision scenarios are games in which each player makes decisions that maximize their own benefits while anticipating the rational choices of their fellow participants (attackers) \cite{rathore2022review}. In game theory, the participant makes decisions and takes actions to produce the desired, ideal result.  In order to obtain the best result of improved network security, SD-SG security studies use gaming theory to evaluate threat scenarios and use the control and data planes as players.
    
\end{itemize}

\subsubsection{Game Theory}Sivaraman et al. \cite{sivaraman2020game} propose a game-theoretic approach to improving data privacy in SD-SG, with the goal of reducing passive information leakage through compromised controllers. The proposed privacy framework is based on the formulation of a noncooperative game among the switches. The requirements for privacy are quantified using information theory mutual information and differential privacy. An iterative best response algorithm is used to compute the game's Nash equilibrium~\cite{daskalakis2009complexity}. The proposed scheme's performance is compared to the globally optimal solution and the exponential mechanism (for differential privacy).When compared to the global solutions, the theoretical game performed nearly optimally on the IEEE 30, 118, and 300 \cite{pstca2018} bus systems. In practice, the proposed approach can be used to improve data privacy in SG systems and protect against compromised controllers.

\begin{figure*}[h]
\centering
\includegraphics[width=1.5\columnwidth]{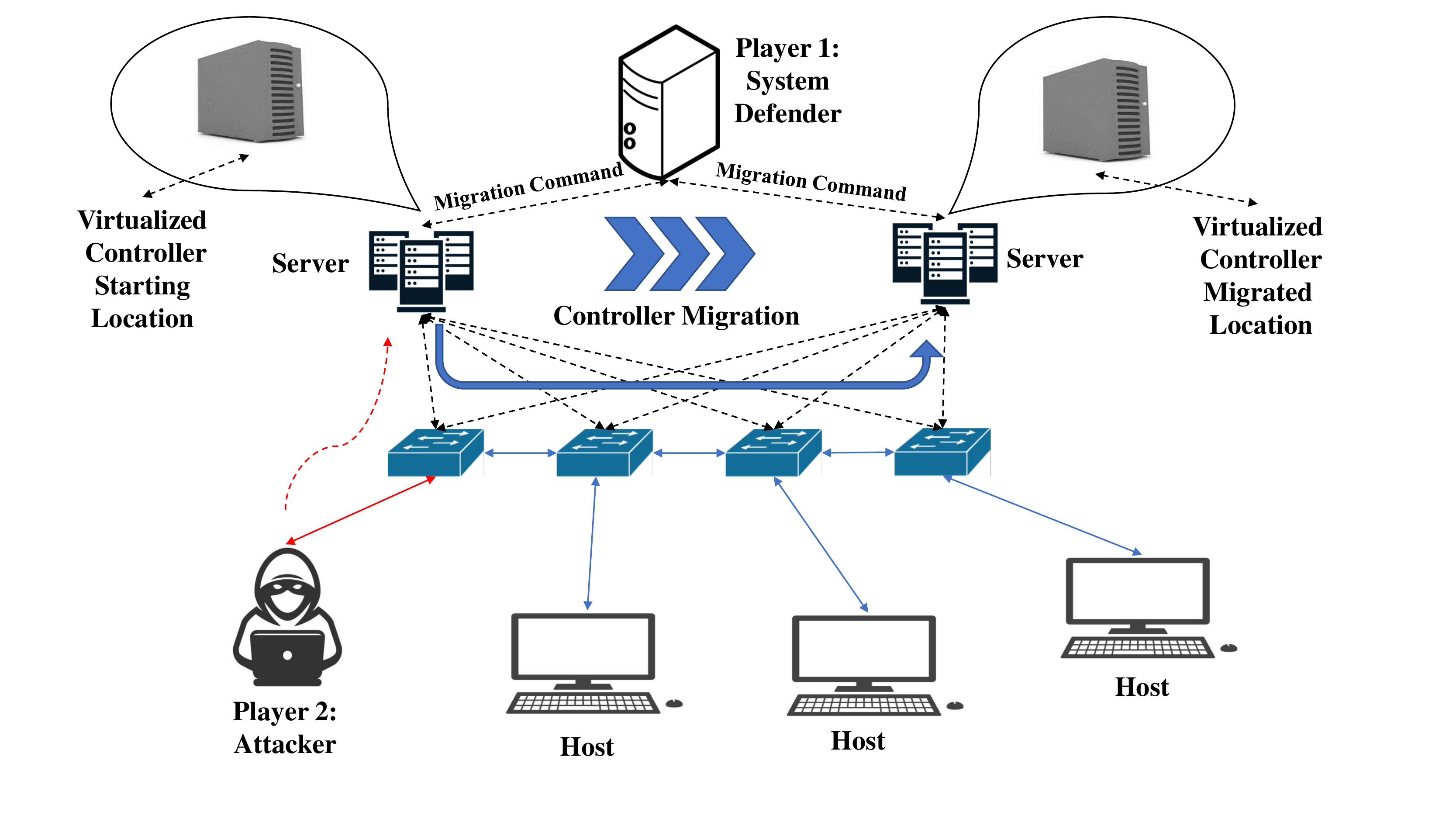}
\caption{SD-SD MTD Game Example}
\label{fig:SD-CPC Game}
\end{figure*}

Samir et al.  \cite{samir2021sd}  suggest a Software-Defined Controller Placement Camouflage (SD-CPC), a stochastic game-based MTD method for improving SDN controller resilience against cyber threats. Figure ~\ref{fig:SD-CPC Game} shows a model of a game-based MTD. The technique seeks to make SDN controllers less vulnerable to attackers by dynamically relocating virtualized controllers and changing their IP addresses. There are two participants in the game: player 1 is the system defender, and player 2 is the attacker. The attacker targeted the most vulnerable regions of the network in their simulations, while the system defender determined the best location to migrate the controllers in reaction. The technique performed particularly well against attacks that tried to exploit vulnerabilities in the controller's software or targeted the controller's location. The proposed game had minmal influence on system performance, and the proposed SD-CPC approach offers an effective and efficient MTD solution for improving SDN resilience to advanced persistent threats to controllers.

\subsubsection{Moving-Target Defense}

 Lin et al. \cite{lin2019security} proposes a Moving Target Defense (MTD) approach based on virtual security functions (VSF) (e.g. firewalls, IDS, traffic classifier, etc.) to improve the security of SDN networks in SG. Furthermore, researchers suggest mitigrating the VSF to resource-rich servers will help mitigate the effects on the VSF. The approach aims to defend against controller attacks by proposing a moving target defense of SD-SG based on a proposed security function virtualization (SFV), making it difficult for attackers to locate and exploit vulnerabilities. The authors used a simulation of a real-world SG topology to assess the effectiveness of the proposed approach. According to the findings, SFV-based MTD can reduce the success rate of various attacks. The method worked especially well against attacks that targeted the controller's location or attempted to exploit vulnerabilities in the controller's software.  Their proposed algorithm achieves the maximum available resource increment, whereas other popular algorithms, random and greedy algorithms, achieve lesser performance. The proposed approach, SFV-based MTD, offers a practical and cost-effective solution for improving the security of SDN-enabled SG. 

Azab et al. \cite{azab2022mystify} proposes "MystifY," a proactive Moving-Target Defense (MTD) approach for improving the resilience of Software Defined Control Plane in Software Defined Cyber-Physical Systems (SD-CPS) against attacks. The strategy aims to defend against advanced persistent threats by changing the controller's network addresses and/or ports on a regular basis, making it difficult for attackers to track the controller's location and exploit vulnerabilities. The authors implemented MystifY on a testbed and assessed its efficacy against various types of attacks. According to the findings, MystifY can reduce the success rate of various attacks by not allowing the attacker to identify the location or identify of the controller. The method worked especially well against attacks that targeted the controller's network location or attempted to exploit vulnerabilities in the controller's software. Furthermore, the authors assessed the proposed approach's impact on SD-CPS performance in terms of network latency, packet loss, and throughput. MystifY had a negligible impact on system performance. Finally, the proposed approach, MystifY, provides an effective and efficient proactive MTD solution for improving the SD-CPS's resilience to advanced persistent threats. The approach achieves a high level of security while having little impact on system performance.

\subsubsection{Final Thoughts for Controller Attack Defense} Controllers, which manage the network and send flows to forwarding devices, are an essential component of SDN frameworks. To prevent cyber attacks, SD-SG network operators must devise methods to secure controllers. To improve the resilience of SDN networks against advanced persistent threats, several moving target defense (MTD) approaches, such as game-theoretic and stochastic game-based methods, have been proposed. MTD approaches involve regularly changing the controller's network addresses and/or ports to make it difficult for attackers to track its location and exploit vulnerabilities. By moving VSF to resource-rich servers, MTD approaches based on virtual security functions (VSF) can also mitigate the effects of attacks.

\subsection{Intrusion Attack Defense}
\begin{figure*}[h]
\centering
\includegraphics[width=2\columnwidth]{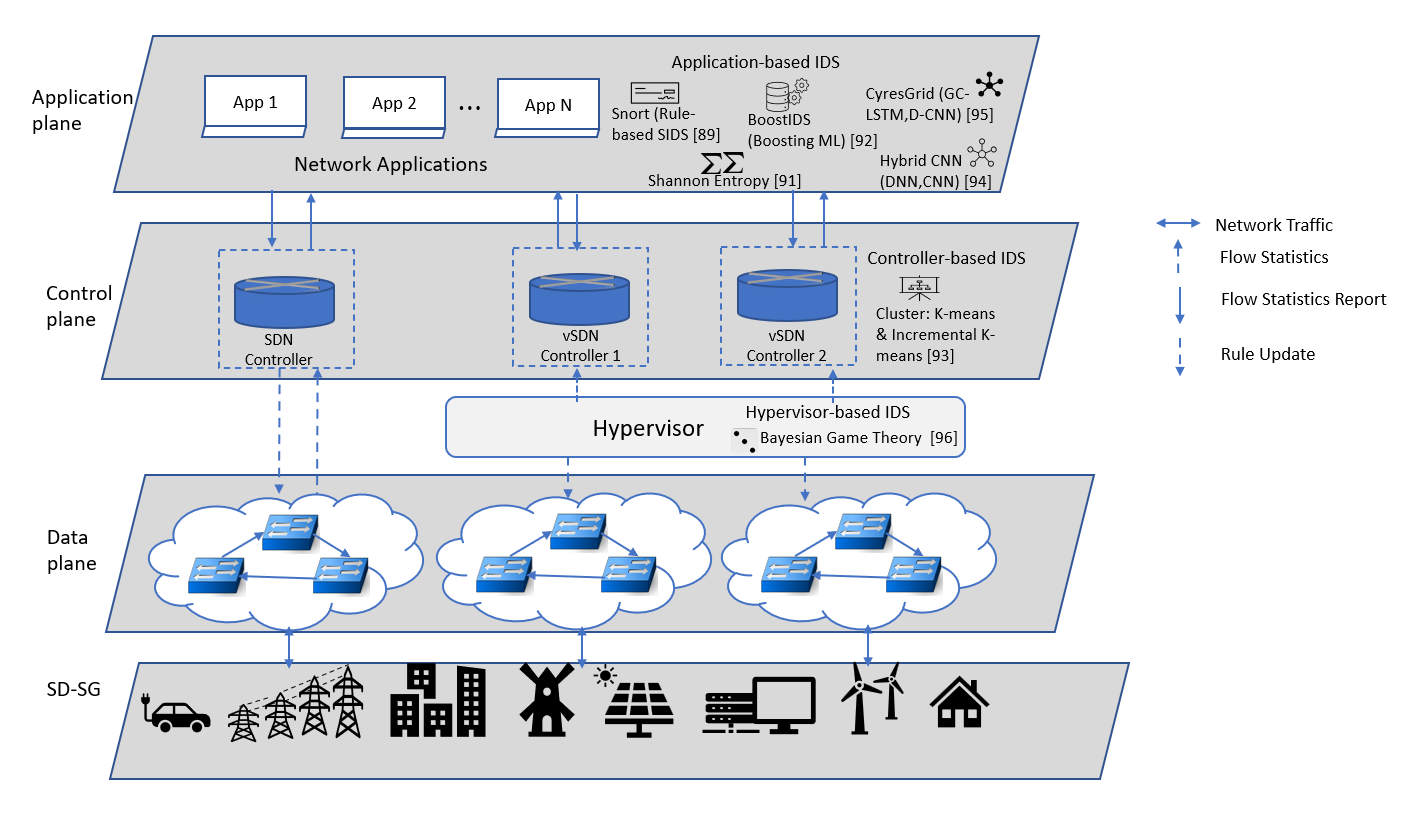}
\caption{IDS Integration in SD-SG Architecture}
\label{IDS_SDSG}
\end{figure*}

\begin{figure}[h]
\centering
\includegraphics[width=\linewidth]{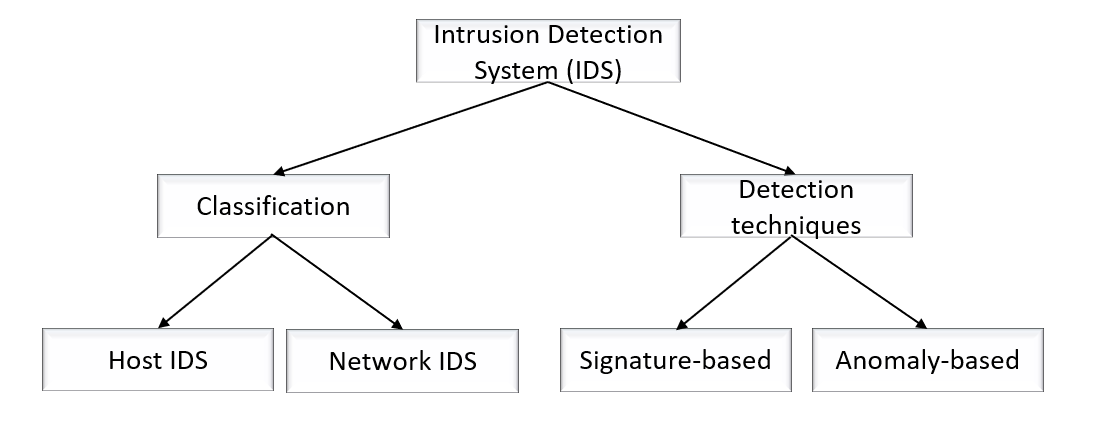}
\caption{Intrusion Detection System Description }
\label{IDS}
\end{figure}

In the ideal scenario, the SD-SG is expected to be accessed through authorized entities to meet the three main cyber-security goals of ensuring confidentiality, integrity, and availability. However, SD-SG vulnerabilities occur as a result of unauthorized access due to intrusion attacks such as false data injection, ARP spoofing, DoS, and host location attacks \cite{rehmani2019software}. For this reason, Intrusion Detection Systems (IDSs) are solutions that detect and report these attacks.
Figure~\ref{IDS_SDSG} illustrates the integration of IDS as an application residing in the SDN architecture. The SDN controller receives flow statistics from the OpenFlow switch and sends them to the IDS, which uses a data processing technique for anomaly detection. The generated report on the traffic flow statistics is sent from the IDS to the SDN controller, which updates the flow rule in the switch to either drop or forward packets based on the received report. IDS can also reside either in the controller or hypervisor and this has been demonstrated in \cite{10.1007/978-3-030-02931-9_7},\cite{8746112}.
IDS as shown in figure~\ref{IDS} is mainly classified as Host IDS (HIDS) and Network IDS (NIDS). HIDS runs on a network device to observe the behavior of other network devices while the NIDS analyzes network traffic \cite{BHARDWAJ2022100580}. NIDS contributes to analyzing SDN traffic flow to identify intrusion attacks. Hence, this section focuses on the proposed detection techniques in NIDS.
Based on how data is processed for attack detection, IDS can be further categorized as Signature-based and Anomaly-based. The Signature-based IDS (SIDS) detects known intrusion attacks using known traffic patterns. Although it is able to correctly detect known attacks with a low false positive rate, it is unable to classify unknown attacks and has a high false alarm rate. In the Anomaly-based IDS (AIDS), it detects intrusion attacks through deviations from normal behavior. Hence, it is able to detect unknown abnormalities. The proposed approaches for implementing SIDS and AIDS as defense strategies against intrusion attacks are categorized and discussed as follows.
\begin{itemize}
    \item \textbf{\emph{Rule-based SIDS:}} SIDS utilizes the rule-based technique which comprises conditional if-then rules to enable predefined signatures to identify known intrusion attacks \cite{9210368}. These rules are easy to create and lightweight. Snort is an open-source rule-based SIDS. It operates by matching packets with the signature database and if there is a rule match, Snort generates an intrusion attack alert \cite{8365396}.
\end{itemize}
\begin{itemize}
    \item \textbf{\emph{Machine Learning:}} Machine Learning (ML) techniques are used in AIDS  for detecting intrusion attacks. It utilizes mathematical models to extract useful patterns from data sets. ML algorithms can be classified as Supervised or Unsupervised. Supervised ML uses labeled data for training to map an input variable to a target variable while Unsupervised ML identifies patterns from unlabelled data by grouping them based on similarities \cite{iasc.2023.026769}. Examples of ML algorithms include K-Nearest Neighbor, Support Vector Machine, K-mean clustering, and Ensemble methods. 
\end{itemize}

\begin{itemize}
 \item \textbf{\emph{Deep Learning:}} Deep Learning (DL) is based on artificial neural networks and uses multiple hidden layers to extract high-level features from data sets. Deep learning has great feature learning capabilities and has recently been proposed to be efficiently applicable for IDS \cite{Peng},\cite{10017381}. DL can be supervised or unsupervised. Examples of DL algorithms include Recurrent Neural Networks, Convolutional Neural Networks, and Deep Neural Networks.
 \end{itemize}

 \begin{itemize}
 \item \textbf{\emph{Game Theory:}} Game theory uses mathematical models for strategic interactions among players such that payoffs are interdependent on each other's actions in instances like attack-defense scenarios. Also, other SDN-related problems like SDN controller assignment, and intrusion detection and mitigation can be modeled as games \cite{8746112}.

 \end{itemize}

\subsubsection{Rule-based SIDS}
Presently, zero-day intrusion attacks are continually on the rise, and the Signature-based IDS is unable to detect these attacks since it uses the classical supervised machine learning techniques or rule-based methods which can only determine similarities in known attacks \cite{Khraisat2019SurveyOI}. This is demonstrated in \cite{info10030106}, where Pedro et. al proposes a DDoS attack detection and mitigation technique that integrates an IDS with an SDN architecture. The authors implement the Snort IDS, which uses known rule-based signatures to detect intrusion attacks namely DDoS, DDoS with IP spoofing, and DDoS with IP packet size manipulation. The proposed method operates in three phases which are the detection, communication, and mitigation phases. The Snort IDS detects DDoS attacks in the detection phase and alerts the SDN controller about the detected intrusion attack in the communication phase. In the mitigation phase, the SDN controller installs some flow rules in the switch to block the intrusion attack. The experimental results indicate that implementing the Snort IDS based on the SDN paradigm mitigates DDoS attacks while maintaining the normal operation of a network. However, the authors used a single Ryu controller which makes the entire network susceptible to a single point of failure. Moreover, the Snort IDS is a Signature-based IDS that detects known intrusion attacks correctly but has a high false alarm rate in identifying unknown intrusion attacks and mutations in known attacks. A potential solution that has attracted research attention is the use of Game Theory, ML, and DL algorithms for Anomalous-based IDS \cite{9664737}. 

\subsubsection{Machine Learning}
Detecting intrusion attacks in the SD-SG is very crucial. Due to this, researchers have put in much work to address this problem by proposing Anomalous-based IDS solutions in defense against intrusion attacks. In \cite{jung2019anomaly}, the authors assert that Entropy-based Anomaly Detection methods that make use of feature distributions of the network have been successful in detecting intrusion attacks like DoS. On this account, they propose the Shannon Entropy to identify anomalies such as DoS and network scanning intrusion attacks in the communication network of an SG using SDN traffic feature distributions like source Internet Protocol (IP) address and destination IP address. A high entropy implies scattering of the feature distribution whereas a low entropy indicates convergence of the feature distribution. Principal Component Analysis (PCA) is used for pre-processing the traffic flow before classification. However, this is ongoing research and the authors expect to implement this proposed entropy-based anomaly detection technique on an SDN-based testbed. Thus, this proposed method has not been validated to confirm its efficiency in effectively detecting intrusion attacks in SD-SG.

According to the analysis conducted in \cite{9843645}, IDS solutions based on a single machine learning model encounter issues like poor generalization and ineffective detection of all attack types. Zakaria et al. design BoostIDS which is a novel framework that leverages ensemble learning to efficiently detect and mitigate security threats like DDoS, probe, fuzzers, and backdoor attacks in SD-SG. BootIDS is deployed as an application in the application plane of the SDN architecture as shown in Figure~\ref{IDS_SDSG} and consists of two modules. The first module uses Boosting Feature Selection algorithm to select relevant SG features. The second module uses a Lightweight Boosting algorithm to effectively detect intrusion attacks in an SD-SG. The experimental results prove that the BoostIDS has higher precision, accuracy, detection rate, and f1-score when compared with existing machine learning intrusion detection systems.

Allen et al. \cite{10.1007/978-3-030-02931-9_7} propose a Hybrid, Distributed, and Decentralized (HDD) SDN architecture to secure the Phasor Measurement Unit (PMU) subsystem network. HDD-SDN utilizes the physically distributed controller approach for fault tolerance and fail-over operations while employing the parallel execution of machine learning models to detect anomalous behavior for a resilient SD-SG. The proposed method uses the parent-child multiple controllers model. The parent controller re-configures the network and allocates and manages resources whereas the child controller determines anomalies in packets and monitors the status of the parent node and other devices within the sub-region network. The authors implement the K-means algorithm in parallel with the incremental K-means algorithm. The standard K-means algorithm recalculates the cluster centers using the entire data set while the incremental K-means updates the previous centers with only newly input data in each iteration. These clustering techniques detect anomalies in the network traffic data and PMU measurement data. The proposed anomaly detection technique using clustering provides approximately 90\%\ accuracy.

\subsubsection{Deep Learning}
A Hybrid Convolutional Neural Network (HYBRID-CNN) is proposed in \cite{Peng} to identify abnormal flow due to intrusion attacks such as scan, DoS, Root to Local (R2L), probe, and User to Root (U2R) attacks in SD-SG. The proposed method uses a Deep Neural Network (DNN) to memorize global features whereas the Convolutional Neural Network generalizes local features for better feature learning capabilities. Although the HYBRID-CNN effectively detects abnormal data flow with a good detection rate, it is biased toward some attack classes. This is due to the unbalanced data sets used in the implementation. However, it provides better results than the compared traditional and deep learning methods. 

Alfan et al. in \cite{10017381} propose CyResGrid, an approach that uses a hybrid deep learning model of a Graph Convolutional Long-Short Term Memory (GC-LSTM) and a deep convolutional network for time-series classification-based anomaly detection in Operational Technology (OT) communication networks for power grids. GC-LSTM comprises two machine learning models, namely, Graph Convolutional Network (GCN) and LSTM. The GCN processes the OT network topological information in the spatial domain. The LSTM learns the time-series data of the observed OT network traffic in the temporal domain. Hence, GC-LSTM is able to learn from both the spatial and temporal domains. The deep convolutional network in CyResGrid uses Bayesian optimization for hyperparameter tuning to detect anomalies. The authors simulated the power system in real-time with the Root Mean Square (RMS) dynamic model of the IEEE 39-bus test system in DIgSILENT PowerFactory whereas the OT network emulation is based on Mininet. They considered DDoS and active reconnaissance which involves OT network scanning as intrusion attacks. The experimental analysis indicates that the proposed CyResGrid outperforms the compared state-of-the-art deep learning-based time-series classification of intrusion attacks. CyResGrid can be improved to detect more variations of intrusion attacks in the SD-SG.

\begin{figure}[h]
\centering
\includegraphics[width=\linewidth]{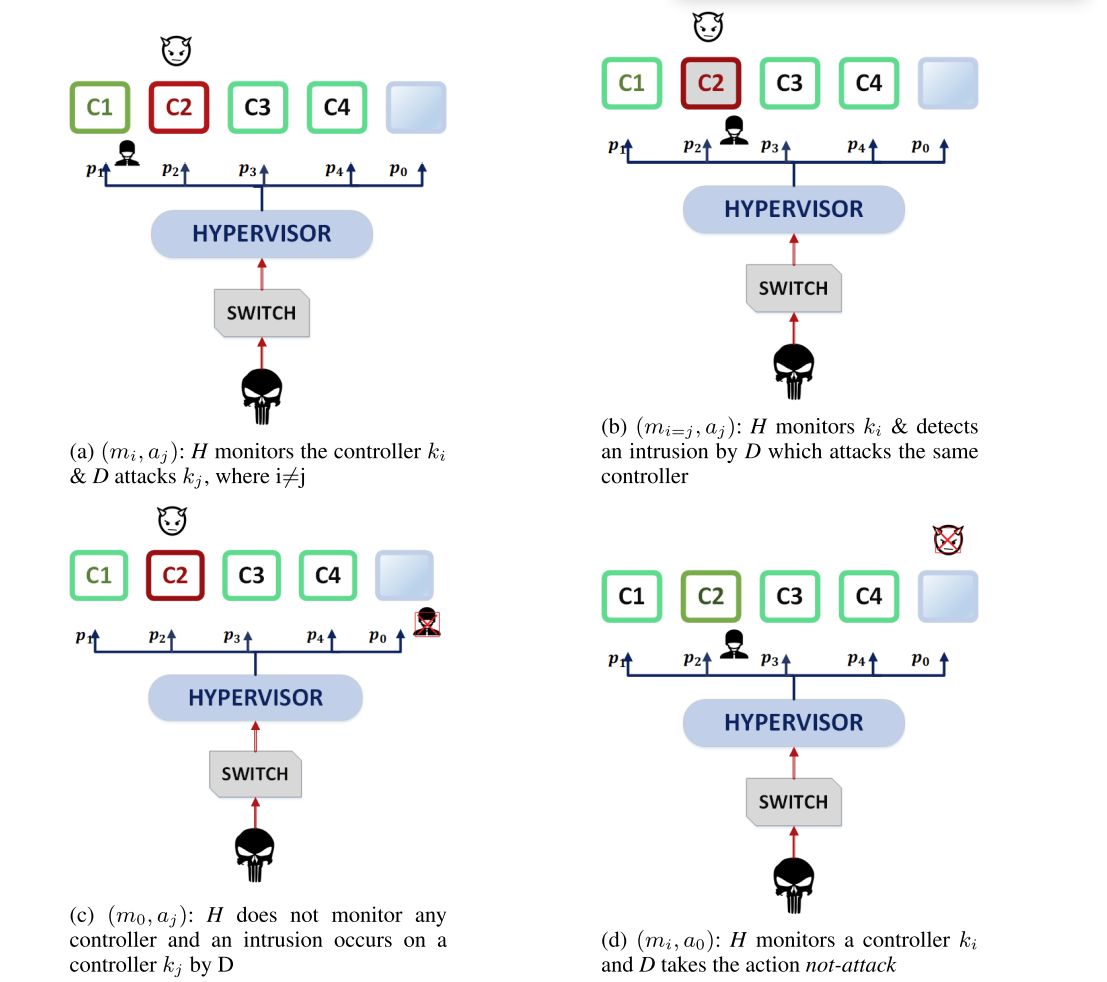}
\caption{Scenarios for Strategic Interactions between a Hypervisor and an Attacker \cite{8746112}}
\label{hypervisor_attacker}
\end{figure}

\subsubsection{Game Theory}
Although the virtualized SDNs (vSDNs) provide better network management and programmability, they are associated with security problems that affect the normal operation of the SG. Rumaisa et al. formulate the strategic interaction between a hypervisor monitoring its vSDN controllers and a possible attack source launching DDoS intrusion attacks via compromised switches as a non-cooperative dynamic Bayesian game-theoretic IDS \cite{8746112}. Figure~\ref{hypervisor_attacker} illustrates the four scenarios considered in the paper for strategic interactions between a hypervisor and an attacker launching DDoS through compromised switches. The proposed game model enables a hypervisor to distribute its limited resources to optimally monitor guest virtualized SDN controllers. In addition, the model provides accurate detection by addressing the realistic approach of a malicious entity, which via a compromised switch aims to minimize its detection by deviating its behavior between a normal and a malicious entity. The simulation results indicate that the non-cooperative dynamic Bayesian game-theoretic IDS enables a hypervisor to increase the probability of detecting distributed attacks, minimizes false positives, and reduces monitoring costs since resources allocations to monitor virtualized SDN controllers depend upon its belief about the source of the attacks that it forms based on its observation. 
\begin{figure*}[!t]
\centering
\includegraphics[width=.9\paperwidth]{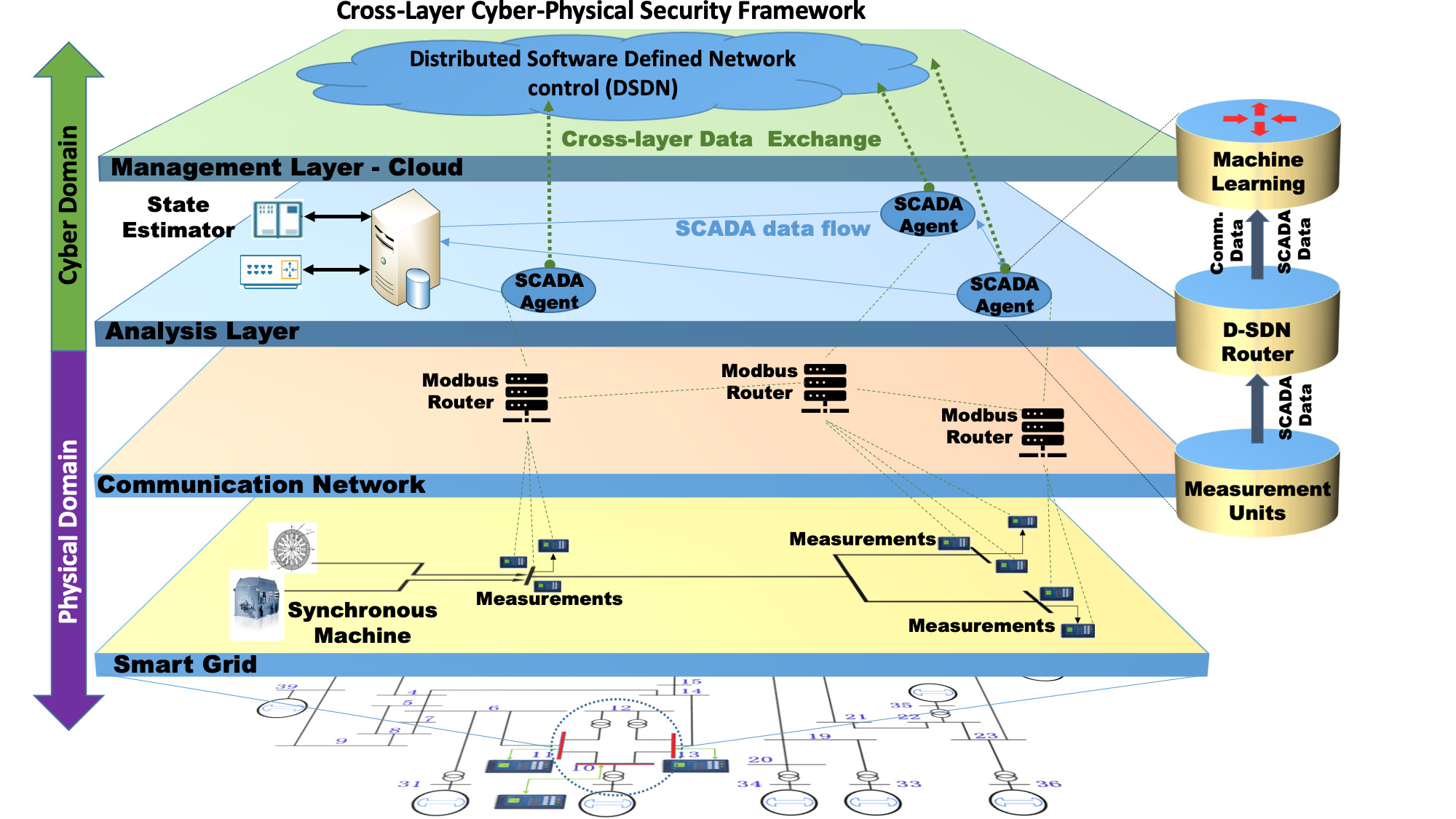}
\caption{Cross-Layer Cyber-Physical Security Architecture Presented in \cite{starke2022cross}  }
\label{fig:Cross-Layer Cyber-Physica}
\end{figure*}

\subsubsection{Final Thoughts for Intrusion Attack Defense}
Intrusion attacks in SD-SG are becoming more complex and their mutation feature makes it difficult to identify them using intrusion detection systems like the Signature-based IDS. Moving forward, current research addresses intrusion attacks by using the Anomalous-based IDS which leverages techniques like machine learning, deep learning, and graph theory. Most recent works are geared toward the use of deep learning algorithms in IDS since these methods have proven to be more efficient in identifying different intrusion attacks in the SD-SG. Moreover, ensemble learning and hybrid models are utilized as well to improve performances in detecting intrusion attacks compared to a single ineffective learning model. A common challenge experienced in the aforementioned literature is the unavailability and imbalance of data sets to implement and evaluate an IDS in the SD-SG. This makes it difficult for IDS models to effectively detect intrusion attacks.

\subsection{Multi-Attack Defense}
\label{Multi-Attack Defense}
At the University of Florida, our previous works \cite{starke2022cross,agnew2022implementation,aljohanicross,nagaraj2020ensemble,trevizan2019data,ruben2020hybrid} have developed security frameworks for SD-SGs. Our publications \cite{starke2022cross,agnew2022implementation } have focused on the detection of multiple cyberattacks using the proposed Cross-Layer Ensemble CorrDet with Adaptive Statistics (CECD-AS) strategy to provide a robust, comprehensive framework that can detect various cyberattack threats such as FDI, DoS, and MITM attacks. In this section, we provide further discussion of the CECD-AS framework and performance analysis as well as survey our work literature that expands upon this framework.

\subsubsection{Cross-Layered Machine Learning Approach}  Allen et al. \cite{starke2022cross} present a cross-layered approach that aims at protecting SGs against cyber threats by incorporating data from the power grid and communication layers for ML model training and testing to safeguard the grid from multiple cyberattacks such as FDI, DoS, and MITM attacks. An overview of the proposed architecture can be seen in Figure~\ref{fig:Cross-Layer Cyber-Physica}. It utilizes data-driven ML techniques such as the CECD-AS algorithm to analyze data from the power grid and SDN-communication layers and detect any anomalous behavior in real-time in the SD-SG as shown in Procedure~\ref{algo::ecd_algorithm}. The equations necessary for the implementation of the CECD-AS algorithm are as follows:

The Mahalanobis Distance Equation: 
\begin{equation} \label{ECorrDet-algorithm} \delta_{m}^{ECD}(\mathbf{z}_{m})={(\mathbf{z}_{m}-\mathbf{\mu}_{m})}^T \Sigma_{m}^{-1} (\mathbf{z}_{m}-\mathbf{\mu}_{m}) \end{equation}

The threshold equation for the CECD-AS Algorithm:
\begin{equation} \label{eq:threshold_noadapt} 
\mathbf{\tau}_m = \mathbf{\mu_{thr,m}} + \eta * \mathbf{\sigma_{thr,m}} 
\end{equation}

Woodbury Matrix Identity \cite{alveyzarecook2016} in equations \eqref{woodbury_mean_ECD} and \eqref{woodbury_cov_ECD} respectively are used: 
\begin{equation} \label{woodbury_mean_ECD} \mathbf{\mu}_{new,m} = (1-\alpha)\mathbf{\mu}_m + \alpha(\mathbf{z}_m-\mathbf{\mu}_m) \end{equation}

\begin{equation} \label{woodbury_cov_ECD} \Sigma^{-1}_{new, m} = \frac{1}{1-\alpha} \left[\Sigma^{-1}_m-\frac{(\mathbf{z}_m-\mathbf{\mu}_m)(\mathbf{z}_m-\mathbf{\mu}_m)^T}{\frac{1-\alpha}{\alpha}+(\mathbf{z}_m-\mathbf{\mu}_m)^T(\mathbf{z}_m-\mathbf{\mu}_m)}\right] \end{equation}

 As discussed in \cite{starke2022cross}, the threshold value $\tau_{m}$ for each local Cross-layer CorrDet detector is updated using equation \eqref{eq:adap_threshold} with updated $\mathbf{\mu}_{thr,m, -\beta}$ and $\sigma_{thr,m, -\beta}$, where $-\beta$ signifies the use of past $\beta$ number of samples for updating threshold

\begin{equation} \label{eq:adap_threshold} 
\tau_{m} = \mu_{thr,m, -\beta} + \eta * \sigma_{thr,m, -\beta}.
\end{equation}

The framework is also designed to enable an efficient integrated approach across various components of the SG. The authors demonstrate the effectiveness of the framework using a case study with results showing greater than $>98\%$ classification accuracy against the aforementioned multiple cyberattacks. Overall, the proposed approach provides a comprehensive and proactive approach to SG cybersecurity, which is essential for ensuring the reliability and resilience of modern power systems. CECD-AS is a continuation of our previous works and ML algorithms \cite{nagaraj2020ensemble,trevizan2019data,ruben2020hybrid} which are discussed next. 

\begin{algorithm}
   \caption{Cross-Layer Ensemble CorrDet with Adaptive Statistics (CECD-AS) algorithm} \label{algo::ecd_algorithm}
    \begin{algorithmic}[1]
    	\State  \emph{ Train a Cross-Layer Ensemble CorrDet classifier}: 
      \Require $\mathbf{Z}$, $\mathbf{Y}$, $\mathbf{\widetilde{Z}}$
      \For {Every local  Cross-layer CorrDet classifier $m = 1:M$} 

      \State Initialize the mean $\mathbf{\mu}_m$ and covariance $\Sigma^{-1}_m$ of normal statistics using the sample mean and covariance of normal samples in the training set with selected \textit{triple elements} associated with $\phi_m$
      \State Initialize the squared Mahalanobis distance $\mathbf{\delta}_{Z,m}$ using equation \eqref{ECorrDet-algorithm}
      \State Initialize the threshold $\tau_m$ using equation \eqref{eq:threshold_noadapt}
            
      \EndFor
      \item[]
          	\State  \emph{Test using the Cross-Layer Ensemble CorrDet classifier with Adaptive Statistics}:
      \item[]
      \For {Every test sample $k = 1: K_2$}
      
      \State Compute the squared Mahalanobis distance $\mathbf{\delta}_{\widetilde{z}_k}$ using equation \eqref{ECorrDet-algorithm}
      
      \If {$\forall m,  \mathbf{\delta}_{\widetilde{z}_k} < \tau_m$}
      \State Classify $\widetilde{z}_k$ as normal sample: $\widetilde{y}_k = 0$
      \State Update the mean $\mathbf{\mu}_m$ and covariance $\Sigma^{-1}_m$ using equation \eqref{woodbury_mean_ECD} and equation \eqref{woodbury_cov_ECD}
      \State Update the sliding window by adding $\mathbf{\delta}_{\widetilde{z}_k}$ to $\mathbf{B}$ and removing the oldest value from $\mathbf{B}$.
      \State Update the mean $\mu_{thr,m, -\beta}$ and variance $\sigma_{thr,m, -\beta}$ of squared Mahalanobis distances in the updated sliding window of each local Cross-layer CorrDet detector 
      \State Update the threshold value $\tau_m$ for each local Cross-layer CorrDet detector using equation \eqref{eq:adap_threshold}
      \Else
      \State Classify $\widetilde{z}_k$ as abnormal sample: $\widetilde{y}_k = 1$
      \EndIf
      
     \EndFor
      \Ensure $\mathbf{\widetilde{Y}}$
\end{algorithmic}
\end{algorithm}

Nagaraj et al. \cite{nagaraj2020ensemble} propose the initial Ensemble CorrDet with Adaptive Statistics (ECD-AS) to detect FDI attacks in the IEEE 118-bus system~\cite{dabbagchi1993power}. The paper presents a method for using adaptive statistics to detect bad data in power systems that takes into account the continuously changing state of a power system. The proposed ECD-AS algorithm extends the work of the CorrDet algorithm \cite{trevizan2019data} and the ECD algorithm \cite{ruben2020hybrid}.  ECD-AS can be understood as a collection of CorrDet detectors that capture adaptive statistics for each local CorrDet environment. The data-driven bad data detection technique proposed in this paper employs adaptive mean, adaptive covariance, and adaptive anomaly threshold calculated with a sliding window approach for incoming data to adapt to changes in system state. Extensive experimentation with the hyper-parameters of the ECD-AS process reveals, in the case study on the IEEE 118-bus system, an optimal solution with much superior bad data detection results than the state-of-the-art ML algorithms by achieving results as high as $99.35\%$ accuracy and outperforming the other algorithms in every other metric (i.e. precision, recall, and F1-score). The ECD-AS algorithm and the research efforts presented in \cite{trevizan2019data,ruben2020hybrid} formed the foundation for the CECD-AS algorithm which made multiple cyberattack detection possible.

\begin{figure}[!h]
\centering
\includegraphics[width=.8\linewidth,trim = 0cm 0mm 0cm 0cm]{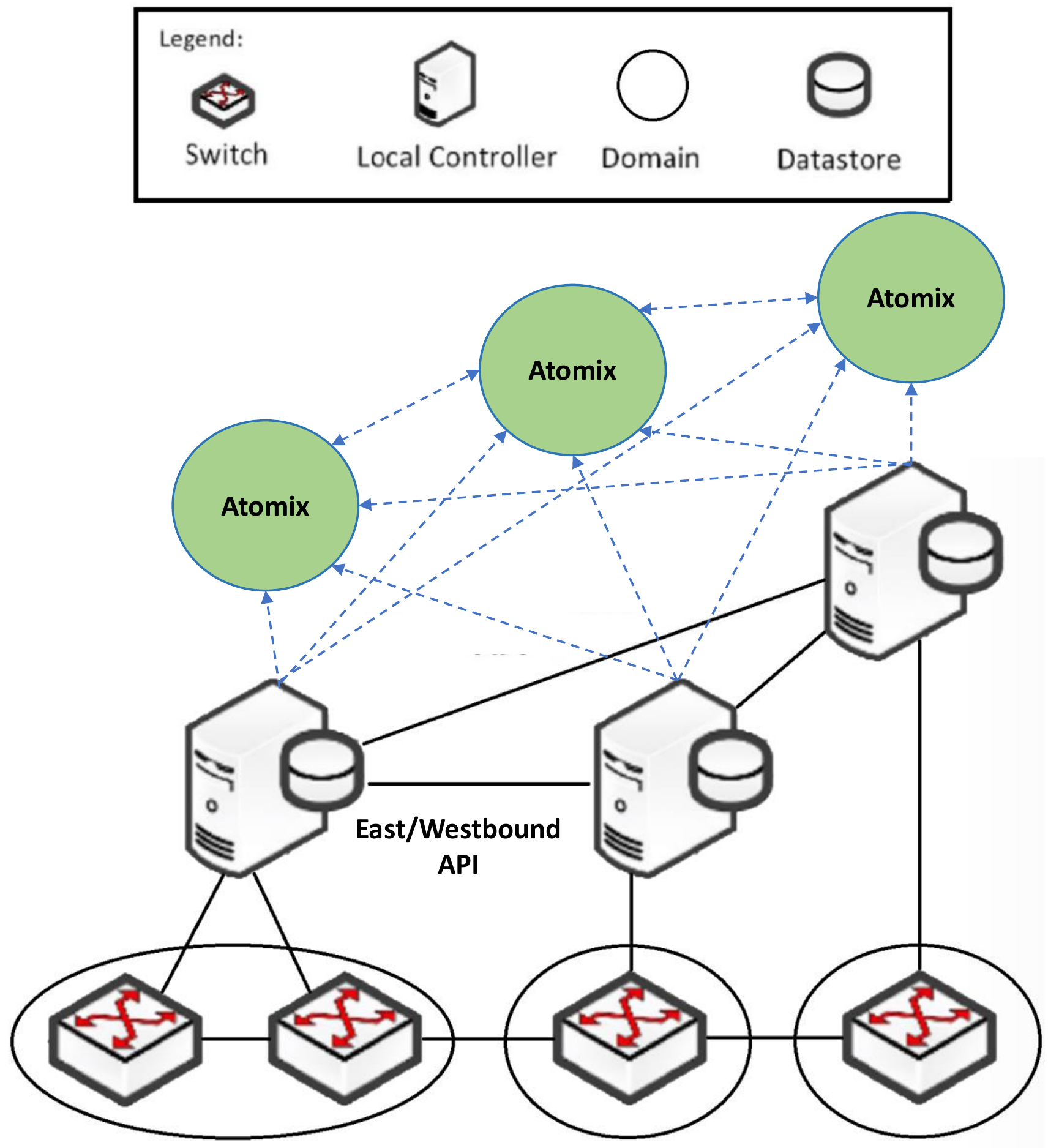} 

\caption{\centering{  Distributed, Flat SDN Controller Architecture \cite{agnew2022distributed,oktian2017distributed}}}
\label{fig:SDN_arch_dist}
\end{figure}

The CECD-AS algorithm was extended in the literature presented by Aljohani  et al. \cite{aljohanicross} which implemented a cross-layered cyber-physical power system state estimation framework that can aid in ensuring grid security based on the CECD-AS approach. To estimate the state of the power system, the framework incorporates data from the physical layer, communication layer and synchronizes the measurements for the CECD-AS ML algorithm. The research demonstrated that a real-time CECD-AS approach outperforms state-of-the-art SE methodologies in terms of F1-score for a variety of cyber attacks by $> 200\%$ due to its ability to learn from data collected in multiple layers of SG and adapt to dynamic spatio-temporal changes in measurement data.

Other works have developed  and extended the SDN architecture layer of the CECD-AS approach to a flat, distributed design for DoS cyberattack resiliency while utilizing the CECD-AS for multiple attack detection. Agnew et al. \cite{agnew2022implementation} proposed a flat, logically distributed SDN communication architecture to improve SG system resilience against cyberattacks. The proposed controller framework can be seen in Figure~\ref{fig:SDN_arch_dist}. To eliminate system of a single point of failure in the control layer, the suggested architecture makes use of three Open Network Operating System (ONOS) ~\cite{berde2014onos} controllers that have distributed control and decision-making control plane for the SD-SG which to as D3-SDN. If one controller goes offline, the other controllers replace and take control of the network for the down controller which significantly improves the SG's resilience to outages in the control layer and load management. A subsequent benchmarking study~\cite{agnew2022distributed} is conducted to compare the performance of the proposed architecture with another common controller solution in SD-SG research, the POX controller\cite{kaur2014network,cokic2019software, mahmood2021s}. The proposed D3-SDN architecture outperforms a single SDN controller framework by increasing throughput by ten times, reducing latency $>20\%$, and increasing throughput during DoS attack situations by approximately $11\%$. 

\subsubsection{Final Thoughts for Multi-Attack Defense}

SD-SG can become subjected to multiple varying cyberattacks such as FDI, MITM, DoS. At the University of Florida, we have developed ML algorithms and cross-layered strategies to detect these attacks with greater performance than other state-of-the-art methods. In addition, we have developed and proposed novel D3-SDN for SD-SG communication layer to mitigate risk to cyberattacks.

\section{Unsolved Challenges}
\label{Unsolved Challenges}
\textcolor{black}{
As SD-SG continues to grow as a critical infrastructure for delivering power and energy to customers, new challenges emerge. One challenge is resiliency, which is defined as how quickly a network can recover victimized nodes and restore connectivity to end users in the event of a cyberattack. DoS attacks pose a significant risk to the resilience of these networks. As previously stated, these kinds of attacks can be conducted to flood target nodes or components in the SD-SG in order to harm or fully shut them down. As stated in Section~\ref{Distributed Denial of Service Attack Defens}, various solutions against DDoS/DoS attacks have been proposed, including blockchain, machine learning, SDN properties, and moving target defenses. Blockchain approaches can increase network security by validating network traffic and the behavior of nodes. However, they require recording, storing, and validating the blockchain process which can add overhead and strain to the system. A lightweight blockchain architecture capable of protecting an SD-SG from DDoS/DoS attacks and providing recovery of nodes has yet to be created. ML models can be taught to detect DDoS/DoS threats, but they fail when it comes to detecting DDoS/DoS attacks launched using zero-day exploits or techniques that the model has yet to be trained and tested on. As a result, trained ML solutions need to be consistently updated with the latest data available which may not be feasible. In addition, a robust ML model that can detect novel DDoS/DoS attacks and provide recovery solutions  for victimized nodes remains to be observed. Other research has focused on using SDN properties such as utilizing the controller's global view of the network to provide detection and mitigation for smurf attacks, sock stress attacks, and SYN flood attacks by employing lightweight Tsallis entropy-based defense mechanisms. Although these tactics are effective for detecting and mitigating these attacks, they do not recover nodes that have gone down because of DDoS/DoS attacks or hijacked nodes used to launch DDoS/DoS attacks. Instead, it focuses on denying port access for identified attacker nodes, which unavoidably causes certain nodes to become unavailable, allowing the DDoS/DoS attack to succeed. The demand for a controller centralization technique capable of recovering fallen nodes remains unfulfilled. Other strategies have used moving target defensive techniques that focus on randomizing link pathways to make DDoS/DoS attacks successful. Although the links are allowed to change on a regular basis, making it more difficult for the attacker to launch a DoS attack, continually changing network topology may result in increased overhead and strain on the controller, lowering overall network performance. A strategy that allows for moving target defense while also providing node recovery has yet to be developed. In summary, these tactics largely focus on detection, prevention, and mitigation rather than recovery of downed nodes, making the DDoS/DoS attack's ultimate purpose effective. The unsolved task of recovering fallen nodes and regaining those nodes with minimal overhead and strain on the SD-SG network infrastructure remains.}

\textcolor{black}{The capacity to protect the privacy of data shared within the SD-SG presents an additional challenge. The privacy risk of these networks is significantly increased by controller attacks. As previously stated, such attacks can be leveraged to take unauthorized control of the SDN controller, which has entire network authority. The SDN controller can be used by attackers to monitor data transmission and, if necessary, divert traffic to hostile hosts.  As mentioned in Section~\ref{controller attack defense}, a number of defenses against controller attacks have been put forth, including moving-target defense and game theory. With the aid of game theory, developers create games in which the security system (a player) must make several decisions in order to achieve a predetermined objective, such as protecting the controller from an attacker (another player). Although game theory can be used to improve controller security and avoid data leakage to ensure data privacy, its implementation inside a system can result in additional system overhead.  Furthermore, a reliable method that can identify both a compromised controller and the attacker who launched the attack has not yet been developed. If the controller and attacker were identified, more effective and precise security measures may be taken. Other studies have concentrated on moving target defense methods such as changing the controllers' various ports to confuse attackers or relocating the controller through the use of software. These techniques can continuously alter the topology, which would be a deterrent to attackers, but they would necessitate regular synchronization of routing logic across the network to ensure proper communication delivery due to continuously fluctuating communication links. Furthermore, it has not yet been realized that a moving target defense can recognize a compromised controller and/or the attacker. In conclusion, these strategies primarily aim to stop controller attacks in order to prevent the controller(s) from being compromised and leaking data. However, the unresolved issue of identifying which controller(s) are compromised, as well as the identity of the attacker, remains unresolved. Therefore, mitigation techniques for IDS techniques remain an unsolved issue.}

\textcolor{black}{It is important for the SD-SG to guarantee reliability in its communication infrastructure by ensuring that very few cyber attacks work with little or no impact using fast recovery mechanisms for sustainable and continuous provisioning of services with no interruptions. However, intrusion attacks, which are launched by intruders through unauthorized access to disrupt the normal operation of the SD-SG make it difficult to provide reliability. Hence, this still remains a major problem in SD-SG, although several literature reviews on counteracting intrusion attacks using IDS have been discussed in Section IV-C. From the discussions, the Signature-based IDS using rule-based methods are efficient in detecting known DDoS intrusion attacks with low response time complexities but are inefficient in detecting new or mutating forms of intrusion attacks. To combat this problem, ML and DL solutions have been explored and applied in IDSs. The Machine learning and Deep learning IDS based on a single model have poor generalization for complex intrusion attack detection. Most research done to resolve this problem involves combining two or more ML and/or DL models, and this has proven to be very effective in detecting intrusion attacks. However, the unavailability and unbalance of real-world datasets strongly influence the performance of ML and/or DL IDS solutions. Moreover, the availability of these datasets and their direct exposure to the wrong hands of malicious users can breach the privacy of these sensitive data and introduce more complex intrusion attacks. To tackle this problem for the performance improvement of ML and/or DL-based IDS to ensure a reliable SD-SG, there is the need to explore more mechanisms that reduce the constraints with training data availability or imbalance, and these methods include transfer learning, federated learning, and data augmentation. The communication network traffic has dynamic patterns, which could be associated with instances such as a change in the network topology or users' service demands. Subsequently, the data training phase of ML and/or DL-based IDS continually has to update its dataset for training to adapt to the dynamic network traffic. Utilizing advanced online training algorithms and automated updates of the ML and/or DL-based IDS depending on the network traffic pattern remains an important research area that can be explored to provide reliability in SD-SG. The reviewed research mostly considers a single SDN controller for updating flow rules, and intrusion attacks on this controller cause single point of failure vulnerabilities resulting in a total shutdown of the communication network. On the other hand, other research works utilize multiple controllers to overcome this issue to ensure fast failovers to sustain normal operations. Although the use of more than one SDN controller improves the reliability of the SD-SG, it presents open research areas such as reducing SDN control overheads, optimally placing and allocating resources to SDN controllers, synchronizing the network states among SDN controllers, and selecting a cluster head from a cluster of SDN controllers. An underlying unsolved problem is that the discussed literature focuses more on intrusion attack detection and further work is required on not only detecting complex intrusion attacks but mitigating these attacks within an optimal time for reliability in the time-critical SD-SGs.} 

\textcolor{black}{As researchers continue to develop unique models and methods to defend SD-SG against attacks, the challenge of adaptability, or defense methods that are not rigidly configured, arises. To genuinely provide a holistic defensive approach against attackers, researchers must examine how flexible their solutions are to numerous attack types and scenarios. Unbeknownst to network defenders, attackers can perform a number of cyberattacks against an SD-SG. A significant portion of current research focuses on specialized SD-SG defense strategies for specific attacks. However, as noted in Section~\ref{Multi-Attack Defense}, there is a new area of SD-SG research focusing on the creation of multi-attack defensive solutions that evaluate data from both physical and network SD-SG components. These solutions primarily employ cross-layered machine learning algorithms that combine cross-layered ensemble learning methods with correlation detectors to dynamically update the statistical model or parameters based on power flow traffic for improved detection of multiple attacks. Cross-layered machine learning algorithms including CECD-AS and ECD-AS incorporate data from both the power grid and the network layer to detect multiple threats (i.e. DDoS, FDI, and MITM) and react to changes in network state. Although these systems can detect many threats, they are data-driven and must be regularly updated with data from various sources. These strategies necessitate data collection from both the power grid and network layers, which may add overhead to SD-SG operations. Furthermore, the model's failure to recognize attacks that it has not seen or trained for remains an unresolved issue. Therefore, the unsolved challenge of a robust defense model for a  multi-cyberattack solution capable of detecting novel or unforeseen attacks has not yet been realized. }


\subsection{Emerging Threats to SD-SG Security}
SD-SG network security techniques have continued to be developed, but attackers are developing new techniques to circumvent them. In the next subsections, we examine emerging potential threats to SD-SG network security in the form of cyberattacks that have yet to be researched for SD-SG and provide potential solutions offered for these attacks in other SDN security literature. As shown in Figure~\ref{fig:taxonomy_SD-SG_attacks}, the emerging threats to SD-SG network security research are the following: Low Rate Denial-of-Service (LDoS) Attacks, Controller Impersonation Attacks, and Black Hole Attacks. 

\subsubsection{Low Rate Denial-of-Service (LDoS) Attack}
    LDoS attacks are stealthier version of the violent nature of DDoS attacks \cite{zhijun2020low}. LDoS attacks send low-rate traffic to a target device or network over a long period of time. LDoS attacks consume network bandwidth, computing power, and memory to degrade the target's performance or availability. LDoS attacks are intended to avoid detection and avoid flooding the target system or network with traffic. LDoS attacks can avoid intrusion detection systems and other security measures that spot high-volume attacks by sending low-volume traffic via short burst of packets over time. Over time this can degrade the QoS of a node while avoiding detection. 
    
    Currently, defense methods of LDoS attacks involve filtering LDoS attacks, improving network parameters, and reallocation of resources \cite{zhijun2020low}. Researchers used the comb filter to filter LDoS attacks from transmission control protocol (TCP) traffic by analyzing the amplitude spectrum to determine periodic parameters of the LDoS attack \cite{wu2017low}. Other Researchers have employed using the random early detection (RED) algorithm which is an active queue management (AQM) method deployed on the router. The RED algorithm avoids congestion by pre-emptively dropping packets before the router's bufer becomes full. Research efforts \cite{zhang2010rred, kuzmanovic2003low} have deployed modified variations of RED and AQM methods on routers to detect LDoS attacks. Other efforts \cite{liu2011q} have deployed machine learning models based on Q-learning to dynamically allocate resources as needed. Imtegration of these algorithms and methods could find vitality in SD-SG network security research.

    
    \subsubsection{Controller Botnet Attacks} 
\begin{figure}[h]
\centering
\includegraphics[width=\linewidth]{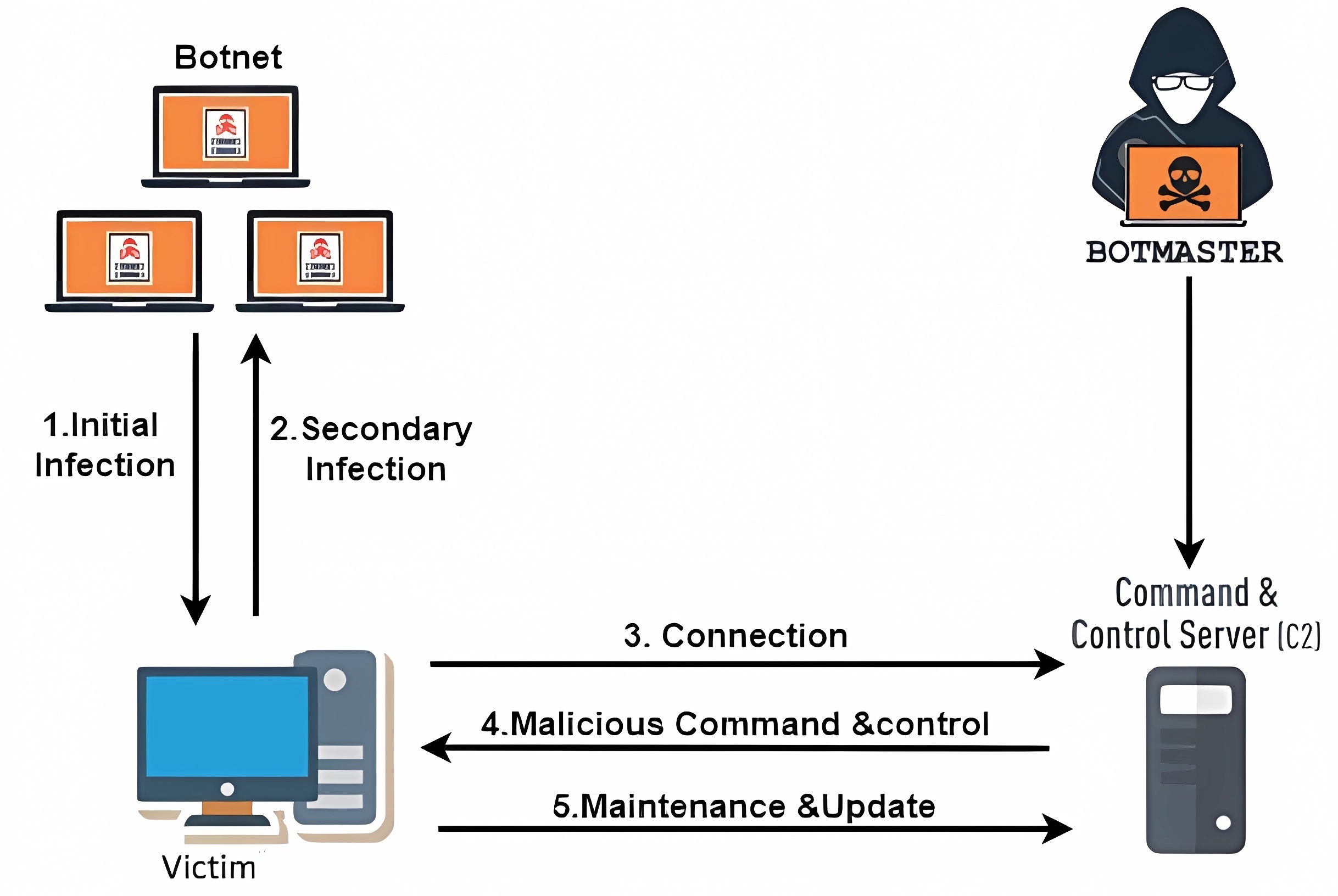}
\caption{Botnet Life Cycle presented in \cite{shinan2021machine}}
\label{fig:botnet_arch}
\end{figure}

Botnets are malware-infected zombie networks that are controlled by a single master called the botmaster. A botnet is comprised of the following elements: bots, botmaster and command and control channel (C\&C)~\cite{shinan2021machine}. Bots are malware-infected computers  that compose the botnet network and can number in the hundreds of thousands. The C\&C channel is a server that is responsible for disseminating commands and receiving information for later access by the botmaster.  With control of the botnet, the botmaster can initiate cyberattacks, disseminate spam or malware, conduct ransom attacks, steal personal information, etc. and cause millions in damages \cite{shinan2021machine}. A botnet architecture is shown in Figure~\ref{fig:botnet_arch}. Additionally, botmasters may offer infected computers to other hackers for use in their own attacks. Because of their crucial role in network governance as aforementioned, SDN controllers are desirable targets for botmasters wants to take control of them for their own advantage. 
    
    Current research has used a variety of methods to detect botnets such as honeypots, IDSs, and machine learning \cite{shinan2021machine}. Honeypots act as decoy nodes on a network for attackers to target and defense frameworks have been developed that use them \cite{luo2019using,ja2021intelligent, shafi2019ddos}. When attackers target this device, this allows for the network operators to gain information of the size of strength of the botnet without exposing the legitimate network nodes \cite{wang2019sdn}. Other research has developed methods that use detect botnet C\&C server bots by using Domain Name System (DNS) queries \cite{sanjeetha2021detection,zafar2019botnet, zha2019botsifter}. The botnet C\&C servers are typically Dynamic DNS (DDNS) providers. Futhermore, researchers have created IDSs to detect botnets \cite{ieracitano2020novel,nguyen2020genetic,ashraf2021intrusion}. With the correct modifications, these techniques could be utilized in SD-SG defense frameworks. 

\subsubsection{Controller Impersonation Attacks} Controller Impersonation attacks involve an attacker transmitting false signals to a device or network in order to mimic a legitimate controller \cite{mutaher2021security, li2016survey}. As a result, the intruder can gain access to and authority within an SDN network without authorization. In order to gain access to sensitive information, alter network traffic, or carry out other attacks, an attacker may pose as a controller and send false signals to SDN-enabled network routers or switches. To address this issue, researchers have proposed authentication models to legitimate controller traffic  \cite{mutaher2021security} or have proposed using IDSs and ML algorithms to detect these attacks \cite{derhab2019blockchain}. Controller impersonation attacks necessitate further research efforts in SD-SG network security because an imposter controller can destabilize the network topology and, as a result, the power grid layer itself. 
    
    
\subsubsection{Black Hole Attacks} A black hole attack is a cyberattack that happens when a malicious node, such as an SDN-enabled router or switch, drops, or 'swallows', every packet it receives, causing a "black hole" in the network \cite{hsieh2018detection, gurung2020survey}. Black hole attacks are one of the most devastating attacks in wireless sensor networks (WSN) \cite{kalkha2019preventing}. WSN networks have been developed and proposed in the current SG paradigm \cite{abujubbeh2019software}.  It is possible that an attacker will take possession of these nodes to make it behave this way. Additionally, the attacker can escalate this attack by seizing control of a controller and forcing nearby switches to direct network messages to the black hole in order to be dropped by modifying the flow tables of the forwarding devices.

Research has proposed a variety of methods for detection of black hole attacks such as IDs~\cite{gruebler2015intrusion, gite2021ml}, Clustering \cite{shi2014cluster, katal2013cluster}, Cryptography \cite{shukla2021mitigate, kumar2021black}, or Trust-based voting schemes \cite{keerthika2019mitigate,naveena2020analysis}. These methods are intended to mitigate, detect, and/or prevent these techniques. Although black hole attacks are more common in wireless networks than in wired networks, SD-SG can use both wired and wireless connections. As a result, it is critical that security for these attacks be developed because a black hole attack can result in data loss and QoS disruption.  

\subsection{Final Thoughts: Unsolved Challenges}
Future SD-SG network security research should concentrate on the development of defenses against the attacks aforementioned. Other SDN applications and research disciplines, such as localized SDN networks, mobile ad hoc networks (MANETs), and software-defined data centers (SDDCs), have developed security frameworks to combat these attacks. However, substantial research and ongoing investigation are deficient in these areas of SD-SG research, which could result in immense harm. If these vulnerabilities are identified in SD-SG's security, cybercriminals could use them to disrupt SD-SG users' QoS. In order to defend the grid from these cyberattacks, SD-SG-specific defense strategies must be developed.


\section{Conclusion}  
\label{conclusion}

Rising electricity consumption, deteriorating infrastructure, and reliability concerns have led to the replacement of traditional power grids with SGs. Current SG systems, on the other hand, necessitate time-consuming manual network management and may encounter interoperability issues due to hardware and software from various vendors. In order to address these concerns, software defined networks (SDN) have been utilized as a method of monitoring and controlling SG communication networks, resulting in software defined SGs (SD-SG) which has improved network administration, visibility, control, and security. However, SD-SG remain vulnerable to cyberattacks that could disrupt power infrastructure operations. As a result, defense and security against these attacks are constantly evolving, necessitating an up-to-date, comprehensive study analyzing these various cyberattacks and defense methods proposed in literature. In this study, we filled this gap for the first time. We characterized and classified current SD-SG network security research efforts, as well as to identify future research needs and open unsolved challenges in SD-SG network security for future research directions.

\bibliographystyle{IEEEtran}
\bibliography{refs}

\end{document}